\documentclass[letterpaper,twocolumn,10pt]{article}
\usepackage{usenix2019_v3}
\usepackage[title]{appendix}
\usepackage{ifthen}

\usepackage{url} 
\makeatletter
\g@addto@macro{\UrlBreaks}{\UrlOrds}
\makeatother

\usepackage{amsthm}

\usepackage{amsmath}
\usepackage{mathtools}
\usepackage{circledsteps}

\usepackage[compact,small]{titlesec}
\titleformat{\paragraph}[runin]{\normalfont\normalsize\bfseries}{}{0pt}{}

\usepackage{enumitem} 
\setitemize{noitemsep,topsep=1pt,parsep=1pt,partopsep=2pt}
\setenumerate{noitemsep,topsep=1pt,parsep=1pt,partopsep=1pt}
\setlist{nolistsep}
\usepackage{subcaption} 

\usepackage{booktabs} 
\usepackage{multirow}

\usepackage{amssymb}
\usepackage{pifont}

\usepackage{listings}
\usepackage{xcolor}
\newcommand{\lmtt}[1]{\fontfamily{lmtt}\fontseries{#1}\selectfont}
\definecolor{darkred}{rgb}{0.75, 0, 0}

\usepackage{tlatex}
\usepackage{color}
\definecolor{boxshade}{gray}{0.85}
\setboolean{shading}{true}

\newcommand{\parabf}[1]{\medskip\noindent\textbf{#1}}

\newcommand{\paraf}[1]{\noindent\textbf{#1}}

\usepackage{xspace}
\newcommand{\sysname}{\textsc{\textrm{Kubedirect}}\xspace}
\begin{document}

\date{}

\title{\Large \bf \sysname: Unleashing the Full Power of the Cluster Manager\\for Serverless Computing}

\author{
{\rm Sheng Qi}\\
Peking University
\and
{\rm Zhiquan Zhang}\\
Peking University
\and
{\rm Xuanzhe Liu}\\
Peking University
\and
{\rm Xin Jin}\\
Peking University
}

\maketitle

\begin{abstract}
FaaS platforms rely on cluster managers like Kubernetes for resource management.
Kubernetes is popular due to its state-centric APIs that decouple the control plane into modular controllers.
However, to scale out a burst of FaaS instances, message passing becomes the primary bottleneck as controllers have to exchange extensive state through the API Server.
Existing solutions opt for a clean-slate redesign of cluster managers, but at the expense of compatibility with existing ecosystem and substantial engineering effort.

We present \sysname, a Kubernetes-based cluster manager for FaaS.
We find that there exists a common \emph{narrow waist} across FaaS platform that allows us to achieve both efficiency and external compatibility.
Our insight is that the sequential structure of the narrow waist obviates the need for a single source of truth, allowing us to bypass the API Server and perform direct message passing for efficiency.
However, our approach introduces a set of ephemeral states across controllers, making it challenging to enforce end-to-end semantics due to the absence of centralized coordination.
\sysname employs a novel state management scheme that leverages the narrow waist as a \emph{hierarchical write-back cache}, ensuring consistency and convergence to the desired state.
\sysname can seamlessly integrate with Kubernetes, adding $\sim$150 LoC per controller.
Experiments show that \sysname reduces serving latency by 26.7$\times$ over Knative, and has similar performance as the state-of-the-art clean-slate platform Dirigent.
\end{abstract}

\section{Introduction}
\label{sec:intro}

Serverless computing, or Function-as-a-Service (FaaS), has gained significant traction in modern cloud applications~\cite{zhang2021faster,wang2018peeking,jegan2023guarding,qi2023halfmoon,lu2024serialization,liu2024jiagu,alzayat2023groundhog,huang2024trenv,szekely2024unifying,stojkovic2024ecofaas,stojkovic2023mxfaas,kuchler2023function}.
Users decompose applications into fine-grained \emph{functions}, and the FaaS platform automates the deployment process~\cite{jia2021nightcore,jia2021boki,kallas2023executing,beldi,zhang2025causalmesh}.
To utilize the cloud infrastructure, FaaS platforms typically rely on low-level cluster managers~\cite{k8s,verma2015large,tang2020twine,schwarzkopf2013omega}.
For example, the Kubernetes cluster manager has given rise to Knative~\cite{knative}, OpenFaaS~\cite{openfaas}, Fission~\cite{fission}, and many others~\cite{kubeless,openwhisk}.

Kubernetes has become the de facto choice for FaaS platforms due to its state-centric principle that enables great extensibility~\cite{burns2016borg,sun2022automatic,sun2024anvil}.
It represents the cluster state as a set of \emph{API objects}~\cite{k8s} stored in etcd, a persistent key-value store~\cite{etcd}. It distributes the cluster management logic across a set of \emph{controllers} that collaborate through a pub-sub service offered by the \emph{API Server} (the etcd frontend).
Controllers subscribe to changes on certain API objects of interest, take actions accordingly (e.g., scheduling), update other API objects to reflect the latest cluster state, and finally publish the updates to the API Server to trigger dependent controllers.
To extend the functionalities of Kubernetes, developers can define custom object APIs and corresponding controllers that translates them to and from the built-in core APIs.
Today, Kubernetes has fostered a rich ecosystem that manages networking, monitoring, storage, and security~\cite{istio,prometheus,redis,terraform}, providing critical services to production-grade systems.

However, extensibility comes at the cost of efficiency due to the \emph{indirect} write-notify round trips to the API Server between controllers.
We find that while controllers are fast with their internal logic (orders of milliseconds), they spend substantial time on \emph{message passing} when exchanging a large amount of state.
Consequently, provisioning hundreds of FaaS instances takes tens of seconds in Kubernetes (\S\ref{sec:background}).
A slow control plane is particularly detrimental to FaaS~\cite{liu2023gap}.
First, massive upscaling is prevalent in FaaS due to its burstiness~\cite{sahraei2023xfaas}, e.g., there can be 50k instance creations per minute in the Azure Functions trace (\S\ref{sec:background}).
Second, upscaling happens on the critical path of function serving, where excess requests are queued until new instances become ready (i.e., ``cold starts''~\cite{yu2024rainbowcake}).
Third, function calls are typically short-lived~\cite{shahrad2020serverless}, making the overhead even more pronounced.
In contrast, creation of the container itself only incurs sub-second or even sub-millisecond latency~\cite{wei2023no,du2020catalyzer,akkus2018sand,oakes2018sock,kohli2024pronghorn}.

Existing solutions opt for a clean-slate redesign of FaaS platforms as opposed to extending existing cluster managers~\cite{singhvi2021atoll,fuerst2023iluvatar,dirigent,zhang2022kole,chen2024yuanrong,szekely2024unifying}.
While these platforms offer more efficient primitives, they fundamentally diverge from the state-centric and modular design philosophy of Kubernetes, which limits their compatibility with the existing ecosystem.
On the one hand, basic functionalities like scheduling and routing, whereas well-developed in Kubernetes, have to be built from scratch.
On the other hand, the cluster state, especially that of FaaS instances, is stored in diverse formats and at different locations across these systems.
Consequently, critical extensions like service mesh~\cite{istio} or monitoring tool~\cite{prometheus} cannot enjoy push-button deployment as they depend on subscriptions to standard Kubernetes APIs, and requires non-trivial manual integration.

In this paper, we show that existing cluster managers can be optimized for FaaS without compromising compatibility.
Our insight is that there exists a \emph{narrow waist} consisting of a few well-known controllers (Figure~\ref{fig:bg:k8s-narrow-waist}) that are always present in Kubernetes-based FaaS platforms.
The narrow waist implements the common functionality of scaling out FaaS instances.
Upstream to the narrow waist are platform-specific controllers that exposes diverse user-facing APIs and configures the narrow waist with custom autoscaling policies.
Downstream are data plane components that route requests and monitor the load based on the output of the narrow waist, i.e., the readiness of FaaS instances.
The fact that the upstream is offline and the downstream read-only implies that (1) performance-wise, the narrow waist is the primary source of cold start latency; (2) implementation-wise, the narrow waist is cleanly separated from the upstream and downstream, allowing for optimizations without affecting external compatibility.

To optimize the scaling process, we bypass the API Server bottleneck with lightweight \emph{direct} message passing through the narrow waist.
Our insight is that the API Server is redundant in two aspects that makes our approach practical.
First, the API Server persists every state transition in the cluster to etcd and enforces exact state recovery across failures.
However, we note that instances halfway through the course of provisioning are naturally \emph{fungible}.
Because these instances are not associated with any physical resources or requests, they are free from side effects and can be safely dropped and replaced across failures.
Moreover, controllers are designed as state machines that continuously query whatever the current cluster state and drive it to the desired one, making them tolerant to rollbacks and capable of recreating the missing instances as needed.
Second, controllers are required to issue updates to the API Server because it can resolve conflicts and serialize concurrent updates to the same object.
However, we find that the scaling process follows a \emph{stage-wise} pattern where controllers sequentially decide the desired state of API objects, allowing for conflict-free collaboration.

A strawman solution of direct message passing is to send the API objects \emph{as is} between controllers.
We observe that there is still a substantial redundancy in existing object APIs that incurs considerable overhead: each controller only updates a few attributes of the objects it manages, e.g., the \emph{Scheduler} only sets the target node, whereas the rest are static and predetermined.
We therefore decouple the dynamic and static attributes and differentiate the APIs used in message passing and internal processing with \emph{dynamic materialization}.
For message passing, the sender only transmits the \emph{delta} attributes for \emph{efficiency}.
For internal processing, the receiver assembles the dynamic and static attributes in-memory, converting to standard API objects for \emph{transparency}.

While the API Server is sub-optimal in terms of performance, it has the benefit of being the single source of truth of the cluster state to all controllers.
\sysname, in contrast, must reconcile the \emph{ephemeral} state introduced by direct message passing across a fleet of decoupled controllers, in the absence of \emph{centralized coordination}, and also correctly enforce end-to-end semantics.
We note that the sequential structure of the narrow waist is analogous to the chain of storage servers in the Chain Replication protocol (CR)~\cite{vanrenesse2004chain}.
However, CR assumes a chain of homogeneous backups that only perform replication, whereas controllers in \sysname are heterogeneous state machines with different state transitions, leading to novel challenges.

First, because controllers can perform non-idempotent operations, e.g., scheduling depends on the varying cluster load, and also in an asynchronous fashion, naively propagating the upstream state like CR can lead to inconsistencies.
Our insight is that we should instead model the narrow waist as a \emph{hierarchical write-back cache}.
Specifically, we opportunistically forward the state transitions of each controller towards the tail, and handle downstream transitions and/or failures as cache invalidations to the upstream.
We design a handshake protocol for joining crash-restarted controllers back to the hierarchy, and lightweight soft invalidation for live controllers.

Second, \sysname should adhere to the conventions on instance lifecycle in Kubernetes, especially for instance termination, i.e., the transition to the \emph{Terminating} state is \emph{irreversible}.
This requirement necessitates separate handling of termination and provisioning, because we cannot invalidate the former like we would the latter.
However, invalidation turns out to be unnecessary because termination is \emph{idempotent}.
We therefore introduce a special type of \emph{Tombstones} objects and perform CR-style replication along the aforementioned opportunistic forwarding pipeline, which implements asynchronous termination, e.g., downscaling.
On top of that, we implement synchronous termination, e.g., preemption, by blocking on downstream invalidations.

We implement a prototype of \sysname on top of Kubernetes and the Knative FaaS platform~\cite{knative}.
We change only $\sim$150 LoC per controller in the narrow waist,
and support external extensions out-of-the-box.
We verify \sysname's convergence to the desired cluster state with TLA+ in the appendix.
In summary, we make the following contributions.
\begin{itemize}[leftmargin=*]
    \item With the design and implementation of \sysname, we affirm that legacy cluster managers can be seamlessly optimized for FaaS without compromising compatibility.
    \item We propose on-demand materialization for direct message passing that achieves both efficiency and transparency. 
    \item We ensure consistency across controllers and convergence like Kubernetes with pairwise state management.
    \item Experiments show that \sysname improves function serving latency by 26.7$\times$ over Knative, and has similar performance as the state-of-the-art system Dirigent.
\end{itemize}


\section{Background and Motivation}
\label{sec:background}

\subsection{Kubernetes Basics}
\label{sec:bg:k8s}

Kubernetes is highly extensible and supports a large ecosystem of extensions.
The CNCF Artifact Hub alone hosts 15.6K extensions~\cite{artifacthub}.
An extension is a set of custom APIs and controllers that collaborate with the Kubernetes core.
Examples include Istio for networking~\cite{istio}, Prometheus for monitoring~\cite{prometheus}, Redis for storage~\cite{redis}, and Jenkins for CI/CD~\cite{jenkins}.

\begin{figure}[t]
\centering
\begin{subfigure}{\linewidth}
\centering
\includegraphics[width=0.8\linewidth]{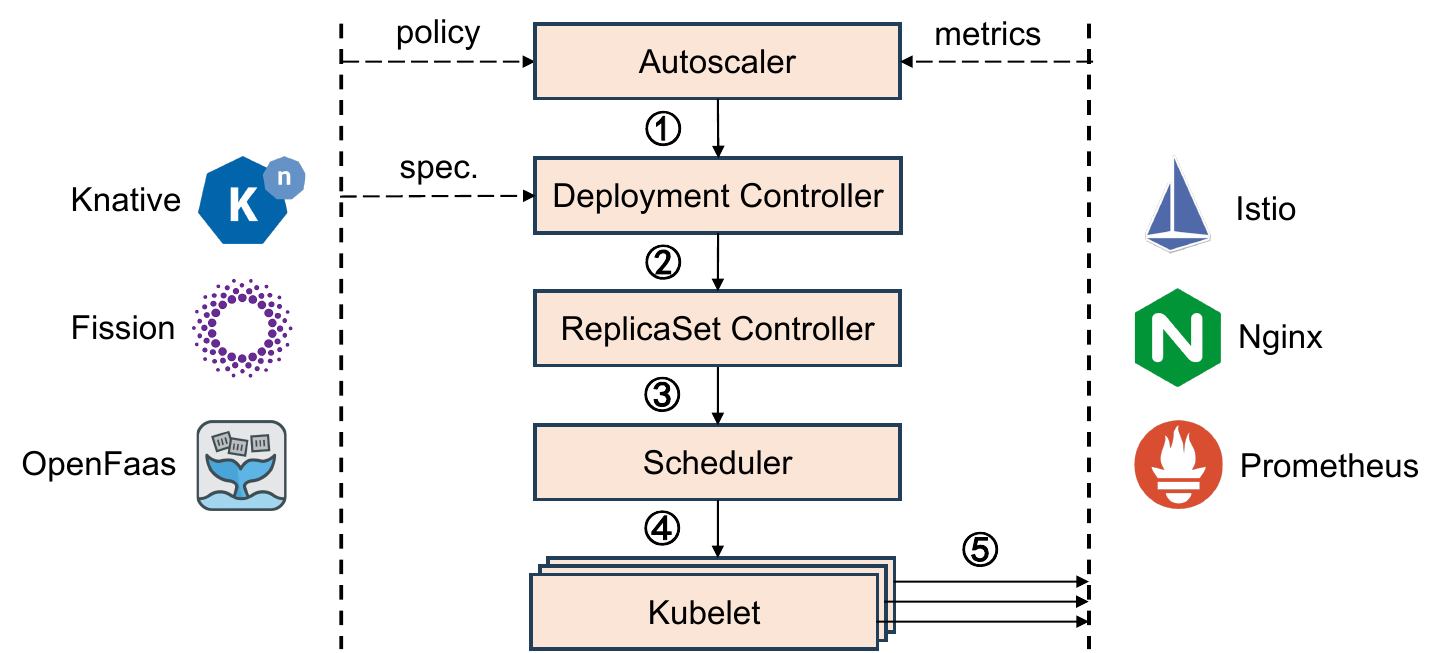}
\end{subfigure}
\hfill
\begin{subfigure}{\linewidth}
\centering
\vspace{1em}
\includegraphics[width=0.9\linewidth]{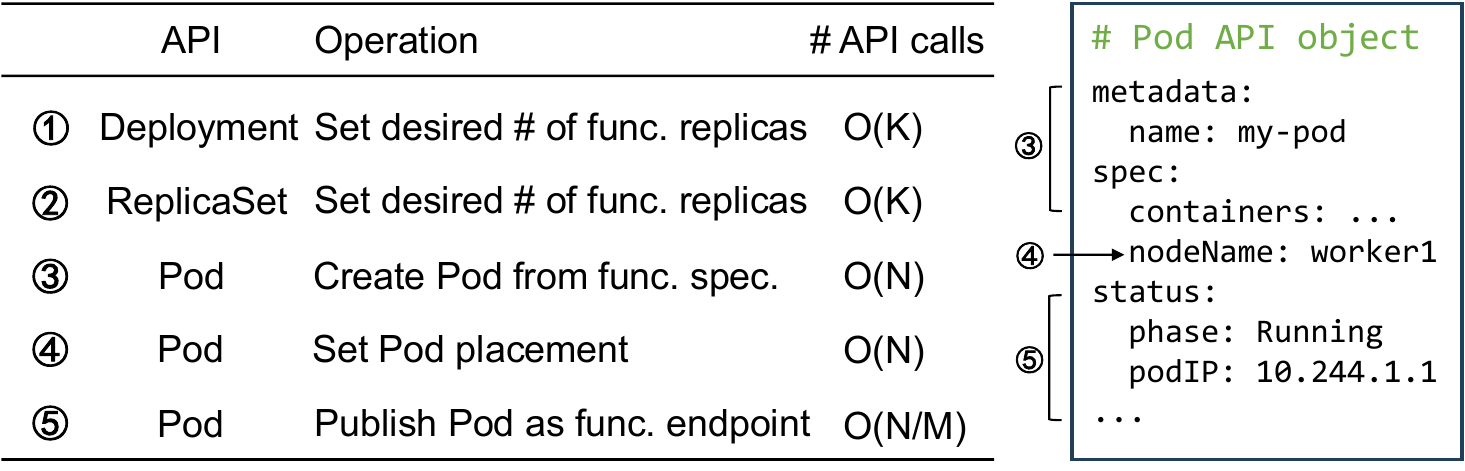}
\end{subfigure}
\vspace{-0.25in}
\caption{The \emph{narrow waist} of Kubernetes-based FaaS platform and the scaling critical path. We consider scaling $N$ Pods for $K$ ReplicaSets in an $M$-node cluster.}
\vspace{-0.1in}
\label{fig:bg:k8s-narrow-waist}
\end{figure}

\begin{figure}[t]
\centering
\includegraphics[width=0.7\linewidth]{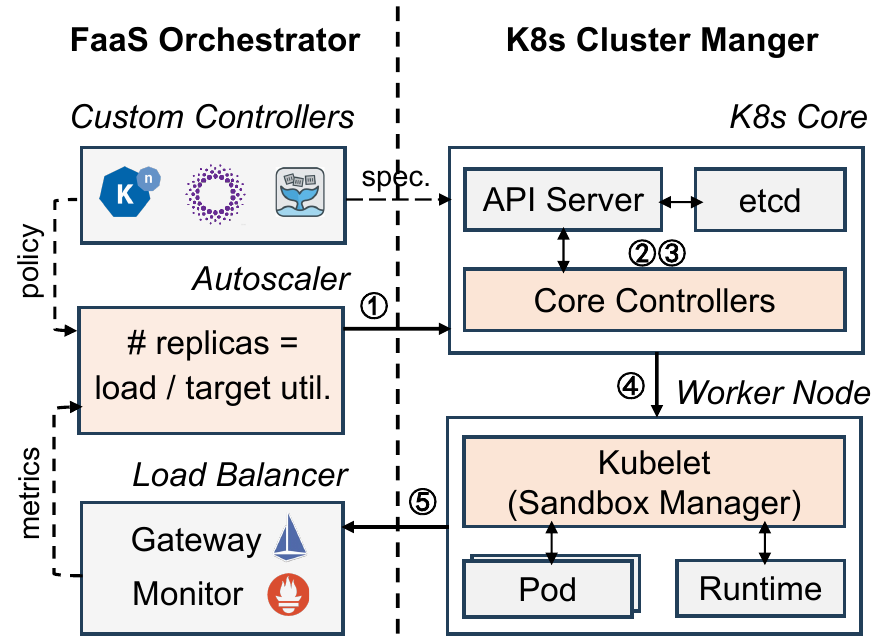}
\vspace{-0.05in}
\caption{Architecture of Kubernetes-based FaaS platform. We highlight the narrow waist in orange. \Circled{1} to \Circled{5} are from Figure~\ref{fig:bg:k8s-narrow-waist} and are indirect calls via the API Server.}
\label{fig:bg:faas-arch}
\vspace{-0.1in}
\end{figure}

The rich ecosystem makes Kubernetes a natural choice for developing FaaS platforms.
We analyze the architecture of three popular Kubernetes-based FaaS platforms: Knative~\cite{knative}, OpenFaaS~\cite{openfaas}, and Fission~\cite{fission}.
Inspite of the diverse user APIs and policies, we find that they share a common \emph{narrow waist} of controllers that implement the basic functionality of scaling out FaaS instances.
Figure~\ref{fig:bg:k8s-narrow-waist} shows the scaling critical path through the narrow waist, and Figure~\ref{fig:bg:faas-arch} shows layout of the narrow waist within the entire platform.

Figure~\ref{fig:bg:k8s-narrow-waist} illustrates the API objects (capitalized) and controllers (italicized) in the narrow waist.
Pod is the basic unit of scheduling, which specifies several containers to serve as one FaaS instance.
ReplicaSet manages a group of Pods that share a common template.
Deployment is a higher-level abstraction of ReplicaSet that implements versioning and rolling updates across versions; it is the Kubernetes-equivalent of a FaaS function.
In case of upscaling, the critical path consists of the following steps.
\Circled{1} The \emph{Autoscaler} computes the desired number of instances based on runtime metrics, e.g. request rate.
It scales a Deployment by updating its {\lmtt{b}replicas} field.
\Circled{2} The \emph{Deployment controller} selects the ReplicaSet of correct version and in turn sets its {\lmtt{b}replicas} field.
\Circled{3} The \emph{ReplicaSet controller} creates a set of new Pods to match the desired scale.
\Circled{4} The \emph{Scheduler} assigns them to cluster nodes by updating the {\lmtt{b}nodeName} field of each Pod.
\Circled{5} The \emph{Kubelet} on each node filters the locally assigned Pods and forward them to the sandbox runtime.
Finally it marks the Pods ready by populating their {\lmtt{b}status} fields and publishes them to downstream controllers outside the narrow waist.

As shown in Figure~\ref{fig:bg:faas-arch}, upstream to the narrow waist are platform-specific controllers that perform offline configurations, and downstream are routing and monitoring components that are read-only to the Pod API.
Therefore, in terms of performance, the narrow waist is the primary source of latency.
In terms of implementation, the differences in upstream configurations are normalized by the common API call \Circled{1} of the \emph{Autoscaler}, and the choices of downstream data plane components by \Circled{5} of the \emph{Kubelet}.
However, the downstream may include standalone services like Istio~\cite{istio}, Nginx~\cite{nginx}, or Prometheus~\cite{prometheus} that, unlike the \emph{Autoscaler}, is not part of the FaaS platform codebase.
Therefore, our prototype of \sysname optimizes \Circled{1} to \Circled{4} and offloads \Circled{5} to the API Server, in favor of compatibility; nevertheless, our techniques are still applicable to \Circled{5}.
Moreover, as we show in the next section, \Circled{5} is not the key bottleneck because it is amortized across all \emph{Kubelets} and nodes in the cluster.

\subsection{The Message Passing Bottleneck}
\label{sec:bg:bottleneck}

While the API Server allows for great extensibility, it complicates the message passing between controllers due to the write-notify indirection.
Previous works~\cite{dirigent,zhang2022kole,cvetkovic2023understanding,apiserver_latency} have identified several factors that make message passing expensive:
(1) the amount of data exchanged, with an average of 17KB per object~\cite{dirigent},
(2) serialization/deserialization, and
(3) persisting to etcd.
Consequently, to avoid overwhelming the API Server and etcd, we find that Kubernetes rate-limits individual controllers in issuing API calls~\cite{why_throttle}, resulting in poor performance when they need to pass a large number of objects to their downstreams.
While simply relaxing the rate limits is known to cause stability issues~\cite{apiserver_latency,etcd_oom}, tuning this configuration turns out to be complicated and labor-intensive~\cite{k8s_tune1k,k8s_tune4k,k8s_tune10k,karthikeyan2023selftune}, with a strong dependency on the actual workload and hardware.
Note that batching API calls will not help in this case, as it simply evades the rate limiter but does not reduce the actual load on the API Server and etcd.
In this paper, we aim to design an efficient message passing mechanism that does not require intricate tuning or fundamental refactoring of Kubernetes's architecture.

\begin{figure}[t]
\centering
\begin{subfigure}{\linewidth}
\centering
\hspace*{0.4em}
\includegraphics[width=0.98\linewidth]{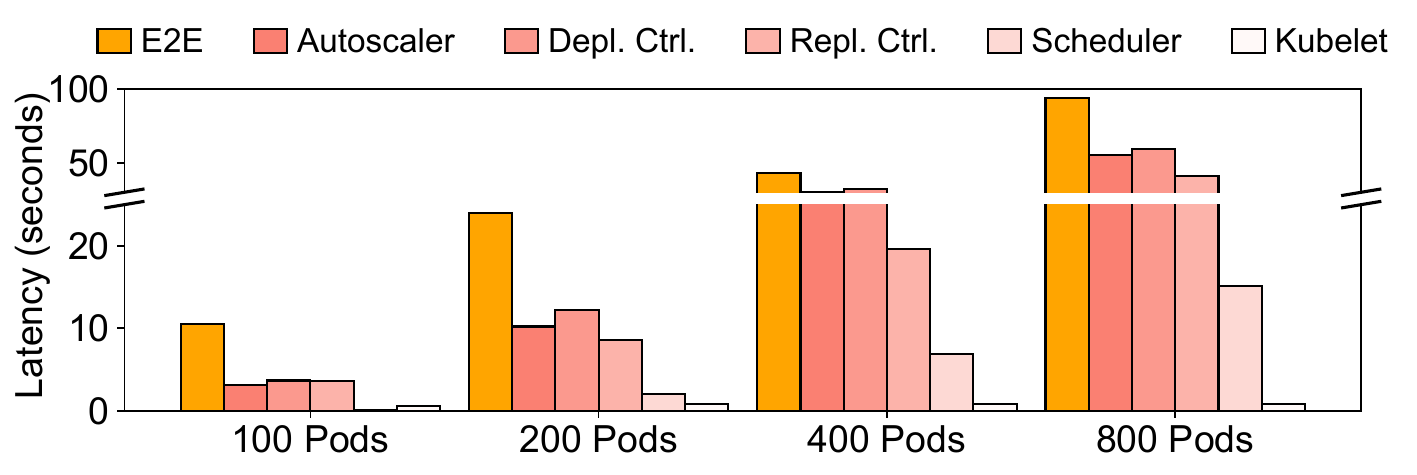}
\vspace{-0.25in}
\caption{The overhead of upscaling.}
\vspace{0.1in}
\label{fig:bg:gap:k8s}
\end{subfigure}
\hfill
\begin{subfigure}{\linewidth}
\centering
\includegraphics[width=\linewidth]{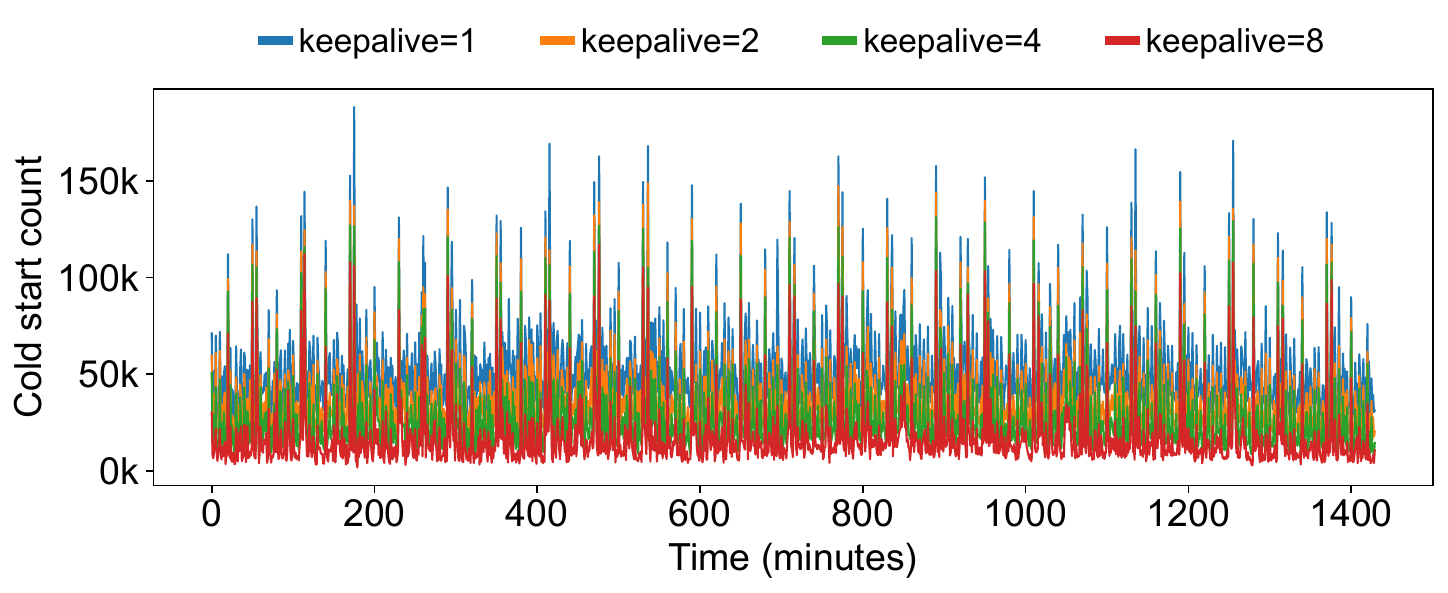}
\vspace{-0.25in}
\caption{The cold start rate in the Azure trace.}
\label{fig:bg:gap:azure}
\end{subfigure}
\vspace{-0.25in}
\caption{The gap between Kubernetes and serverless.}
\label{fig:bg:gap}
\end{figure}

To understand the status quo overhead, we measure the end-to-end (E2E) latency of FaaS upscaling with varying number of Deployments and Pods in an 80-node Kubernetes cluster.
For simplicity, we scale up one Pod per Deployment in the experiment.
To break down the latency, we also measure the time each controller would take if the upstream messages were instantaneous, e.g., if all to-be-scheduled Pods arrive all at once at the \emph{Scheduler}.
Figure~\ref{fig:bg:gap:k8s} shows that controllers in the narrow waist face non-trivial bottlenecks, except the \emph{Kubelets}.
Examining their output logs, we find that controllers are fast with their internal logic (100s of milliseconds), and spend most of the time passing objects to the next controller (\Circled{1}--\Circled{4} in Figure~\ref{fig:bg:k8s-narrow-waist}).
Although they work in a pipelined fashion, the end-to-end latency is still dominated by the slowest stage.
The \emph{Kubelets} are more scalable because they are only responsible for the subset of Pods assigned to the respective nodes (\Circled{5}).
While it is possible to replicate and shard the other controllers in a similar fashion, doing so requires non-trivial refactoring of the Kubernetes codebase and even more tuning, which contradicts our design goal of minimum intrusion and maximum reuse.

The message passing bottleneck via the API Server makes Kubernetes inadequate for serving bursty FaaS workloads where massive scaling is prevalent.
Figure~\ref{fig:bg:gap:azure} shows the cold start rate in the Azure Functions trace over 24 hours~\cite{shahrad2020serverless}.
There can be over 50k cold starts in a minute, far beyond the capability of the standard Kubernetes control plane.

\subsection{Opportunities for Efficient Message Passing}
\label{sec:bg:oppo}

\paraf{Bypassing the API Server}.
We note that traversing the API Server is not a fundamental requirement for the narrow waist.
First, the API Server persists every state transition (to etcd).
However, a closer look at the transitions \Circled{1}--\Circled{4} in Figure~\ref{fig:bg:k8s-narrow-waist} shows that they belong to two categories that both do not require persistence.
(1) \Circled{1}\Circled{2} are \emph{level-triggered}; the desired number of replicas is repeatedly computed in each interation of the autoscaling loop, without needing to memorize the last decision.
(2) \Circled{3}\Circled{4} work with \emph{uninstantiated}, \emph{fungible} Pods that only declare their resource requirements and are free from external side effects, allowing them to be discarded and recreated across failures.
Second, the API Server is required to resolve conflicts and serialize concurrent updates to the same object.
However, the \emph{sequential} structure of the narrow waist allows for conflict-free, one-to-one message passing.
That is, each controller \emph{exclusively} decides the desired state of objects it manages (i.e., one writer), and each object maps to a \emph{single} downstream (i.e., one reader).
For example, although the \emph{Scheduler} is the upstream to all \emph{Kubelets} in the cluster, each Pod is only relevant to its designated \emph{Kubelet}.
Therefore, bypassing the API Server is practical for the narrow waist.

\parabf{Eliminating the redundancy in messages}.
A strawman solution is to send the API objects \emph{as is} between controllers.
However, as described in \S\ref{sec:bg:bottleneck}, prior works also identify the amount of data exchanged and serialization/deserialization as a problem~\cite{dirigent,zhang2022kole,apiserver_latency}.
Quantitatively, we find that it incurs 20\%-35\% overhead (\S\ref{sec:eval:ablation}).
Our key observation is that most API calls in Figure~\ref{fig:bg:k8s-narrow-waist} only update a small subset of object attributes, whereas the rest are static and predetermined, suggesting a substantial redundancy.
For example, when creating a Pod (\Circled{3}), the Pod specification is copied from the template specification from its parent ReplicaSet.
To eliminate the redundancy, we can decouple the dynamic and static attributes and differentiate the APIs used in message passing and control loops, so that we can achieve efficiency for the former and transparency for the latter.

\subsection{Challenges for State Management}
\label{sec:bg:challenge}

The key problem of direct message passing is that it introduces a set of ephemeral objects across controllers.
Unlike Kubernetes, \sysname cannot rely on the API Server as the single source of truth.
Instead, it must provide controllers with a consistent view of the cluster in the absence of \emph{centralized coordination}.
Inspired by the sequential structure of the narrow waist, we note that there is an opportunity to reduce the problem of \emph{global} consensus to \emph{pairwise} agreements, similar to the Chain Replication protocol (CR)~\cite{vanrenesse2004chain}.
CR is designed to manage a chain of storage servers.
It receives update requests at the head and then replicates the update sequence down the chain, providing linearizable reads and writes.
When a server crashes, CR simply removes it from the chain and reconnects adjacent servers.
However, \sysname fundamentally differs from CR as controllers are distinct state machines that can progressively update the state of objects, rather than backups that strictly follow the primary.

We summarize two major challenges.
First, certain non-idempotent controller operation, e.g. Pod scheduling, makes it unsafe to simply propagate the upstream state as in CR (\S\ref{sec:design:problem}).
Therefore, \sysname must judiciously chooses between forwarding or rolling back the state of controllers.
Second, to seamlessly integrate with the ecosystem, \sysname should comply with certain rules of Pod lifecycle.
Specifically, a Pod's transition to the \emph{Terminating} state must be \emph{irreversible}, requiring that termination be handled differently from upscaling, which are free from such constraints.
Moreover, termination can be asynchronous, for overall downscaling, or synchronous, for preemption by high-priority services.
\sysname should handle these nuances effectively while integrating with the upscaling pipeline.


\section{\sysname Overview}
\label{sec:overview}

\begin{figure}[t]
\centering
\includegraphics[width=0.9\linewidth]{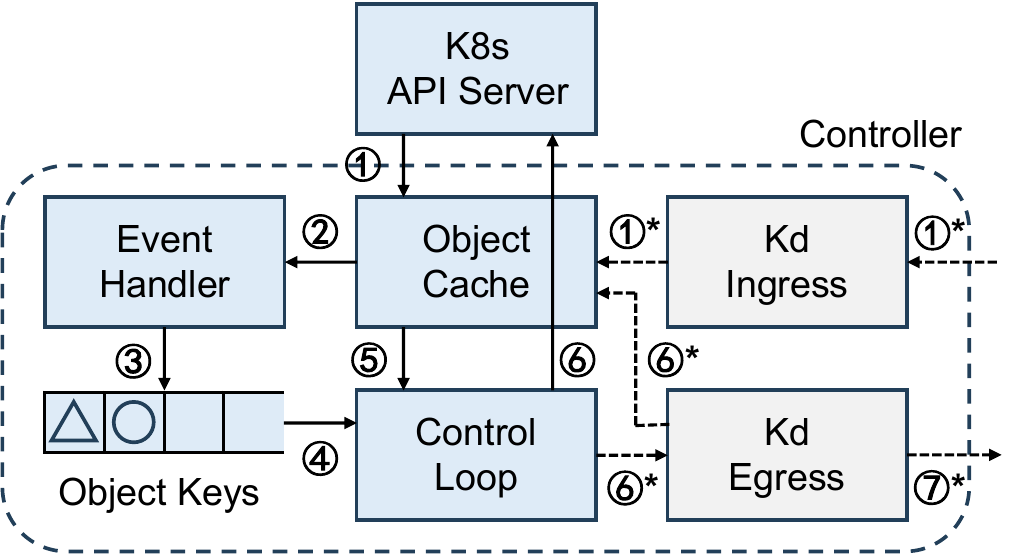}
\vspace{-0.1in}
\caption{\sysname's (Kd) integration with Kubernetes (K8s) controllers. The added steps are labeled with asterisks.}
\vspace{-0.1in}
\label{fig:overview:arch}
\end{figure}

The core of \sysname is a library for exchanging and reconciling the state of controllers in a distributed manner, providing efficiency, transparency, and consistency at the same time.
We build a prototype of \sysname on top of the Knative~\cite{knative} FaaS platform, but our library is applicable to other platforms and essentially to any chain of controllers.

To let \sysname manage the scaling of a FaaS function, users simply add a special annotation to the matching Deployment object; they can switch back to standard Kubernetes by removing the annotation.
\sysname follows the same critical path as in Figure~\ref{fig:bg:k8s-narrow-waist}, but performs direct message passing with efficiency and transparency.

\subsection{Architecture}
\label{sec:overview:arch}

Figure~\ref{fig:overview:arch} shows the integration of \sysname with a standard Kubernetes controller.
We first introduce the basic controller architecture and then discuss how \sysname can be seamlessly integrated.

\parabf{Basic architecture}.
Kubernetes controllers adhere to a \emph{uniform} state-centric architecture~\cite{sun2022automatic,sun2024anvil}.
Specifically, a controller maintains a local cache that \Circled{1} subscribes to the API Server and gets notified on updates to certain API objects of interest.
The notifications also \Circled{2} trigger a set of controller-specific event handlers that decide which objects needs to be reconciled and \Circled{3} push the object keys to a work queue.
The control loop of the controller then \Circled{4} dequeues the keys, \Circled{5} fetches the corresponding objects from the cache, and takes actions accordingly, e.g., creating new Pods or scheduling a Pod to a node.
Finally it exports its updates by \Circled{6} invoking the Kubernetes API.

\parabf{\sysname integration}.
Two adjacent controllers in the narrow waist are connected by a TCP-based bidirectional link.
The downstream link forwards the desired state, while the upstream link sends feedback signals such as cache invalidations (\S\ref{sec:design:cache}).
Specifically, \sysname adds a pair of ingress and egress modules for \sysname-managed, ephemeral objects, parallel to the pub-sub workflow of Kubernetes in Figure~\ref{fig:overview:arch}.
Note that we enforce a clean separation of ownership between \sysname and Kubernetes to prevent any conflicts (\S\ref{sec:impl}).
We integrate \sysname into the basic architecture with the following steps, marked with asterisks; the circled numbers show which steps in the original workflow they are parallel to.
$\Circled{1}^*$ The ingress converts \sysname's messages to standard API objects with dynamic materialization and merges them into the cache, \emph{transparently} triggering the internal control loop.
$\Circled{6}^*$ The egress intercepts the publication of API objects, $\Circled{7}^*$ converts them to \sysname's messages, and sends them to the downstream.
Note that prior to step $\Circled{7}^*$, the egress can immediately populate the local cache with the latest state.
This is because each controller can solely decide the state of the object it manages per the sequential structure of the narrow waist.

\subsection{Dynamic Materialization}
\label{sec:overview:dynmat}

\paraf{Minimal message format}.
\sysname decouples the dynamic and static attributes of API objects and only passes the dynamic ones for efficiency, using the minimal message format in Figure~\ref{fig:overview:msg}.
\emph{Dynamic materialization} is the process of translating to and from standard API objects, providing transparency to the control loop.
Specifically, a \sysname message is set of key-values pairs; a key references a specific attribute of an API object, whereas the value can be an arbitrary literal, representing dynamic attributes, or an external key that points to some static attribute in another object.
For example, Figure~\ref{fig:overview:msg} shows a message from the \emph{Scheduler} to the \emph{Kubelet} representing ``PodX'' on node ``worker1''.
It includes a external pointer to the ``replicasetY'' object. The \emph{Kubelet} can retrieve the ReplicaSet from its local cache, copy its template, and use it to construct the target Pod.

\parabf{Extensibility}.
Because the Kubernetes APIs have a well-defined schema, controllers can use reflection~\cite{go_reflection} to decode \sysname messages such that they remain loosely coupled.
Our handshake protocol (\S\ref{sec:design:cache}) further allows extending the narrow waist with new controllers, e.g., one after the \emph{Scheduler} that sets node-specific variables for Pods.

\parabf{Complexity}.
In terms of the upscaling process in Figure~\ref{fig:bg:k8s-narrow-waist}, \sysname has the same algorithmic complexity as \Circled{1}--\Circled{5}, but significantly reduces the constant factor.
Apart from bypassing the API Server, our minimal message format uses up to 64B per object, whereas the size can be up to 17KB in Kubernetes~\cite{dirigent}.
\sysname can further reduce the message passing overhead by batching messages.

\begin{figure}[t]
\centering
\begin{lstlisting}[numbers=none,morekeywords={KdKey,KdValue,KdMessage}]
// May start with objID in case of external pointer
type KdKey {
    string attrPath
}

type KdValue union {
    string value
    KdKey ptr
}

type KdMessage {
    string objID
    dict[KdKey, KdValue] attrs
}

// Example message from Scheduler to Kubelet
KdMessage Pod {
    objID: "podX",
    attrs: {
        "spec": "replicasetY.spec.template.spec",
        "spec.nodeName": "worker1",
    }
}
\end{lstlisting}
\vspace{-0.1in}
\caption{The minimal message format in \sysname.}
\vspace{-0.1in}
\label{fig:overview:msg}
\end{figure}

\section{State Management}
\label{sec:design}

This section presents the state management of \sysname that maintains consistency across controllers and ensures end-to-end semantics of instance lifecycle.
We start with an analysis of the problem in \S\ref{sec:design:problem} and introduces our mechanisms for instance provisioning in \S\ref{sec:design:cache} and termination in \S\ref{sec:design:lifecycle}. We discuss the correctness of our design in \S\ref{sec:design:proofs}.

\subsection{Problem Analysis}
\label{sec:design:problem}

As per \S\ref{sec:bg:oppo}, the \emph{Autoscaler} and the \emph{Deployment controller} are level-triggered and idempotent, for which fast-forwarding suffices.
The focus of this section is reconciling the state of Pod objects across the \emph{ReplicaSet controller}, the \emph{Scheduler}, and the \emph{Kubelets}, in the context of upscaling.
We discuss downscaling and termination in \S\ref{sec:design:lifecycle}, which have more subtle semantics.
Also note that although it currently takes only three stages of controllers to manage Pods, our analysis and approach applies to arbitrary numbers of sequential stages.

\parabf{Why is fast-forwarding unsafe?}
Inspired by the sequential structure of the narrow waist and the best-effort nature of FaaS scaling, a straightforward solution is to fast-forward the upstream state to the downstream like in Chain Replication (CR)~\cite{vanrenesse2004chain}.
However, controllers are fundamentally different from CR backups because they can update their local state independently or even non-idempotently, rather than strictly replicating the upstream.
When compounded with the asynchrony between the modular controllers, this can lead to unexpected anomalies.

\parabf{Anomaly \#1}.
Consider a \emph{Kubelet} that is temporarily disconnected from the \emph{Scheduler}.
During this period, it evicts a Pod due to temporary resource contention.
However, the \emph{Scheduler} later reconnects with the \emph{Kubelet}, finds the Pod missing, and fast-forwards it again.
The \emph{Kubelet} could publish the terminated Pod again, possibly with a different IP.
This is not allowed in Kubernetes because it violates the convention on instance lifecycle (\S\ref{sec:design:lifecycle}), and could have unexpected effects over external controllers outside the narrow waist.

\parabf{Anomaly \#2}.
Suppose the \emph{Scheduler} restarts after a crash but only reconnects with a subset of \emph{Kubelets} in the cluster due to network partition.
It is possible that a Pod is cached both at the \emph{ReplicaSet controller} (the upstream) and the unreachable \emph{Kubelet} (the downstream), but not at the \emph{Scheduler} itself.
The \emph{ReplicaSet controller}, having not received the placement of that Pod from the \emph{Scheduler} prior to the crash, fast-forwards the Pod again.
The \emph{Scheduler} could in turn assign it to a different node, leading to undefined behavior.

\parabf{Why assume a hierarchical write-back cache?}
We draw an important conclusion from the anomalies: due to the state being mutable through the narrow waist, \emph{the downstream becomes the single source of truth}.
CR assumes the opposite where the head server in the chain acts as the primary.
Therefore, upstream controllers must make decisions based on the downstream state, rather than naively fast-forwarding their own.
It follows that the problem is in fact similar to managing a \emph{hierarchical write-back cache}.
Specifically, the upstream controller opportunistically issues writes to the downstream, which acknowledges the writes but does not guarantee that they will be applied due to active cancellation or passive failures.
Our job is to reflect the downstream changes to the upstream as \emph{cache invalidations}.
We note that our analogy is practical because (1) the sequential structure of the narrow waist implies that the downstream holds all the state the upstream needs to know; (2) controllers are designed to be tolerant to asynchronous notification of events, in which cache invalidations and API Server subscriptions are similar, and will continously attempt to amend whatever the current cluster state to match the desired one.

\subsection{Hierarchical Write-Back Cache}
\label{sec:design:cache}

We build the hierarchical cache on top of the bidirectional links between adjacent controllers as described in \S\ref{sec:overview:arch}.
The upstream propagation goes on the forward link, while downstream invalidations goes on the backward link.
We consider two types of cache invalidations:

\begin{itemize}[leftmargin=*]
    \item \textbf{Hard invalidation} occurs whenever the upstream disconnects and reconnects with the downstream. It \emph{atomically} resets the upstream state to the downstream's before performing any further actions.
    \item \textbf{Soft invalidation} occurs between a live and connected pair of controllers where the downstream \emph{incrementally} informs the upstream of its state changes, e.g., when the \emph{Scheduler} sets the target node for a Pod.
\end{itemize}

In brief, our goal is to use a combination of hard and soft invalidations to ensures a consistent view of the cluster across controllers, in spite of their asynchrony.
As per \S\ref{sec:design:problem}, this guarantee is sufficient for controllers to function correctly.
We defer the proofs of correctness to \S\ref{sec:design:proofs}.

Specifically, the goal of hard invalidation is to provide a consistent initial state across controllers, such that subsequent state exchanges, either forward or backward, can be incremental.
Hard invalidation also provides a uniform solution to controller or network failures, which is desirable because it can be difficult to proactively distinguish between the two cases in distributed systems~\cite{lou2022demystifying,lou2025deriving,huang2018capturing,zhang2019inflection,pan2024efficient,lu2025onesize,wu2024efficient}.

We implement soft invalidation in the same way as forward message passing, employing dynamic materialization (\S\ref{sec:overview:dynmat}).
Next, we present our design of a handshake protocol that implements hard invalidation.

\begin{figure}[t]
\centering
\begin{lstlisting}[moreemph={tcp},morekeywords={connect,accept,send,recv,is_empty,get_state,merge,reconcile}]
def ClientHandshake(cache):
    conn = tcp.connect()
    server_state = conn.recv()
    if cache.is_empty(): # recover mode
        cache.set(server_state)
    else: # reset mode
        change_set = cache.diff(server_state)
        cache.set(server_state)
    # use soft invalidation to propagate change_set
    return conn, change_set

def ServerHandshake(cache):
    conn = tcp.accept()
    serer_state = cache.get_state()
    conn.send(serer_state)
    return conn
\end{lstlisting}
\vspace{-0.1in}
\caption{Pseudocode of the handshake protocol.}
\vspace{-0.1in}
\label{fig:design:handshake:code}
\end{figure}

\parabf{The handshake protocol}.
Figure~\ref{fig:design:handshake:code} shows the pseudocode of the handshake protocol.
The protocol is initiated by the upstream controller (as the client) towards the downstream (as the server).
Once complete, it returns a connection handle for subsequent state exchanges.

Specifically, the downstream controller listens for incoming connections and responds with its local state.
Because it is the source of truth in our hierarchical cache analogy, it can immediately finishes its part of the handshake.

The upstream controller, after receiving the downstream state, can operate in two modes.
(1) In recover mode, the controller has crash-restarted and has empty local state. It therefore applies the downstream state as is.
(2) In reset mode, the controller has non-empty local state and must actively resets it.
There are two cases for each object in its local state. If the object is also present in the downstream, it is simply overwritten and marked dirty. Otherwise, it is marked as invalid but not immediately removed from local state; it is hidden from the internal control loop such that it is equivalent to being deleted.
After completing the handshake, the controller will use soft invalidation to notify its further upstream of the marked objects in either case.
Because soft invalidation is asynchronous and may overlap with the control loop, retaining the set of objects with the invalid mark allows the controller to ignore any incoming updates for those objects.
These objects can be discarded once the further upstream acknowledges them.
Eventually, the marked change set will be propagated to the entire upstream.
Note that this is a guaranteed behavior even though soft invalidations are best-effort; should any upstream controller crash, it would populate its state \emph{``the hard way''}, which would surely incorporate all changes because the downstream source of truth would be guaranteed to do so.

\parabf{Overhead}.
Our approach has no extra overhead compared to Kubernetes, since the API Server also needs to populate or refresh the cache of controller, which can already be more expensive than the dynamic materialization in \sysname.
Apart from that, \sysname also employs another optimization in the handshake protocol: when running in reset mode, the upstream controller only requests the \emph{version number} of the downstream objects in the first round; it fetches only the change set in the second round.
The version numbers do not have to be ordered; they can be any unique numbers because we only care for equivalence.
The key difference from synchronizing with the API Server is that \sysname needs to propagate changes with multiple hops, but this is not a major concern because the narrow waist is shallow.
We demonstrate the efficiency of our approach in \S\ref{sec:eval:micro}.

\begin{figure}[t]
\centering
\begin{subfigure}{0.49\linewidth}
\centering
\hspace*{0.4em}
\includegraphics[width=0.9\linewidth]{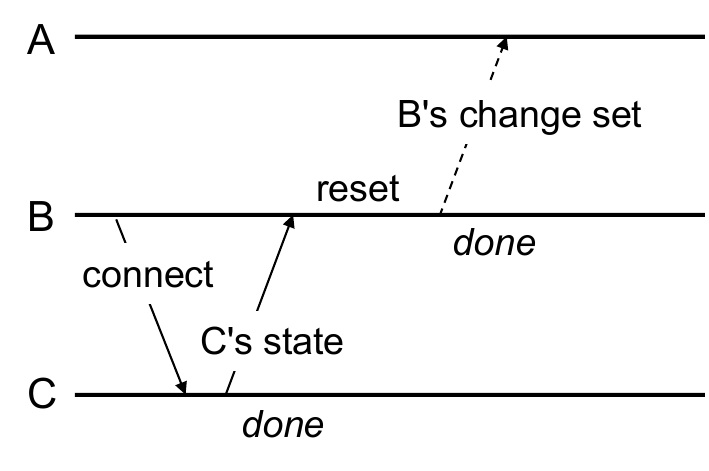}
\vspace{-0.05in}
\caption{Handling B--C disconnection.}
\label{fig:design:handshake:disconn}
\end{subfigure}
\hfill
\begin{subfigure}{0.49\linewidth}
\centering
\includegraphics[width=0.9\linewidth]{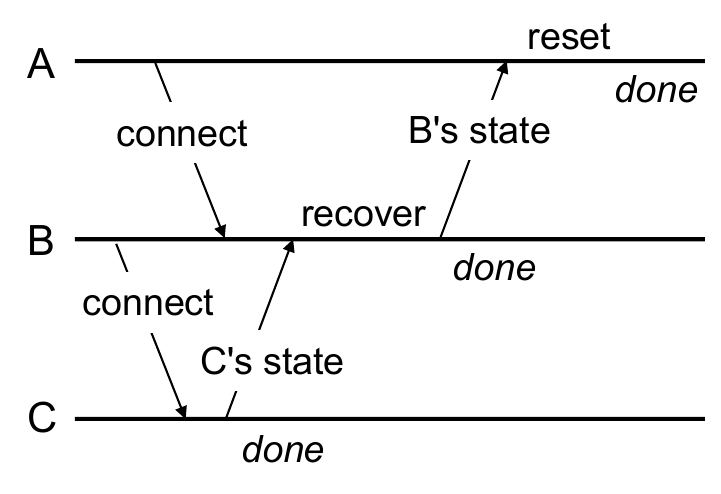}
\vspace{-0.05in}
\caption{Handling B's crash failure.}
\label{fig:design:handshake:crash}
\end{subfigure}
\vspace{-0.1in}
\caption{Failure handling using the handshake protocol.}
\vspace{-0.1in}
\label{fig:design:handshake}
\end{figure}

\parabf{Autonomous recovery}.
The handshake protocol provieds a uniform solution to controller or network failures.
Figure~\ref{fig:design:handshake} shows an example.
Network disconnections can be fixed with a single round of handshake in reset mode.
Controller crash failures can be fixed in two rounds, first with the downstream in recover mode, and then with the upstream in reset mode.
Following the downstream-first rule, the protocol naturally extends to multi-point failures.
Moreover, we can use the protocol to join new controllers into the narrow waist, similar to handling crash failures.

\subsection{Instance Lifecycle Management}
\label{sec:design:lifecycle}

Our design of the hierarchical cache ensures that controllers eventually hold a consistent view of the cluster, in the context of upscaling.
This section takes downscaling into consideration, and also enforces a high-level yet more subtle requirement: the state transitions observed by controllers should adhere to certain conventions on Pod lifecycle.
This is important because \sysname is designed to seamlessly integrate with the Kubernetes ecosystem.

\parabf{State diagram}.
We outline a simplified state diagram of Pod lifecycle.
A Pod starts in the \emph{Pending} state, and transitions to \emph{Running} when it becomes ready. During this process, it may enter the \emph{Terminating} state, which is required to be irreversible in Kubernetes convention.
A \emph{Terminating} Pod will be eventually removed from the cluster state.

\parabf{Provisioning vs. Termination}.
Without loss of generality, we consider two types of events that trigger lifecycle transitions: Pod provisioning and Pod termination.
We identify a fundamental difference in the semantics of the two events that calls for separate handling.
Specifically, Pod provisioning is well-suited to the opportunistic state forwarding and inherently tolerant to sporadic downstream reset, while Pod termination is not.
This is because resetting can have different meanings in the two cases.
For Pod provisioning, resetting leads to the loss of some assumed Pods, which can be considered as the transition from the \emph{Pending} state to the \emph{Terminating} state.
However, for Pod termination, resetting a Pod that is assumed to be \emph{Terminating} could ``revive'' the Pod (e.g., when the decision to terminate is dropped somewhere downstream), which is undefined behavior.

Consequently, it turns out that the upstream \emph{once again holds the source of truth} in the context of \emph{active} Pod termination.
Note that we distinguish \emph{active} termination from the \emph{passive} case, e.g., Pod lost due to node crashing, because the former involves controllers assuming certain transitions should happen, while the latter can simply inform controllers of the status quo.
We also note that terminating a particular Pod is inherently an \emph{idempotent} operation, i.e., it removes an existing Pod or becomes a no-op, 
which adheres to the assumption of the Chain Replication Protocol (CR)~\cite{vanrenesse2004chain}.

\parabf{Replicating Tombstones}.
Based on the above observation, we introduce a special type of API object internal to the narrow waist---the Tombstones.
A Tombstone contains the identifier of a certain Pod, which means that it is marked for \emph{best-effort} termination within the controller's \emph{current session}, which lasts until the controller itself crashes.
During the current session, the controller will continuously replicate the Tombstones over the opportunistic forwarding pipeline.
Tombstones are subject to CR-style fast-forwarding in case controllers crashes or disconnects.
The only possibility of losing a Tombstone is when \emph{every controller the Tombstone has been replicated to crashes}, implying that the termination, whereas best-effort, has a high possibility of success.
A controller can stop replicating a Tombstone when it finds that the Pod it references is locally present but not downstream.
This is well-defined based on our design of the hierarchical cache (\S\ref{sec:design:cache}), and the fact that new Pods are always created at the upstream and propagated downstream.
This controller can then remove the Pod, which in turns triggers invalidation signals to its upstream.
We can garbage collect a Tombstone once the corresponding Pod is removed.

\parabf{Practical examples}.
We demonstrate how to handle actual cases of active termination based on Tombstones. We enumerate the possible scenarios as follows:

\begin{itemize}[leftmargin=*]
    \item \textbf{Downscaling} occurs when the \emph{ReplicaSet controller} receives a smaller desired scale. It selects a set of Pods to terminate, creates Tombstones for them, and starts replicating them downstream. It does so in an \emph{asynchronous} fashion, i.e., it can continue to process subsequent scaling requests, but uses the Tombstones to track the Pods awaiting termination, to avoid unnecessary thrashing.
    \item \textbf{Preemption} can occur at the \emph{Scheduler} or the \emph{Kubelets}. Because the \emph{Kubelets} are at the tail of the narrow waist, Tombstones are unnecessary, and invalidations alone suffice. The \emph{Scheduler} preempts in a similar way as downscaling, except that it must do so in an \emph{synchronous} fashion. This is because there could a high-priority Pod whose placement is conditioned on the termination of the victim Pod. Therefore, the \emph{Scheduler} waits for the invalidation signal from the downstream \emph{Kubelet}.
    \item \textbf{Cancellation} occurs in a special case where the \emph{Scheduler} connects with only a subset of \emph{Kubelets}. The \emph{Scheduler} could be running, in which case it wants to cancel the Pods on the unreachable node; or has crash-restarted, as in \emph{Anomaly \#2} (\S\ref{sec:design:problem}), in which case it does not know the Pods running on the unreachable node, and wants to drain the node to enforce a consistent view of the cluster state. Due to the disconnection, \sysname cannot rely on direct message passing to inform the target \emph{Kubelet} of its decision. Instead, the \emph{Scheduler} marks the corresponding Node API object as invalid through the API Server. The target \emph{Kubelet} is expected to drain all \sysname-managed Pods once it sees the mark. Meanwhile, the \emph{Scheduler} can safely assume that the related Pods are irreversible terminated, and can proceed to send invalidation signals.
\end{itemize}

\parabf{Remark}.
Tombstones are similar in functionality to the invalid marks in the handshake protocol.
However, we decouple the two because they are propagated in the opposite direction, follow different sources of truth, have different lifetimes, and require separate handling in case of failures.

\subsection{Correctness}
\label{sec:design:proofs}

Like Kubernetes, \sysname has the same guaranteed convergence to the desired state, i.e., it ensures that eventually there will be the desired number of Pods running in the cluster.
Because reasoning over such a complicated distributed system involves numerous nuances, we use TLA+ to verify the end-to-end properties of \sysname.
Nevertheless, to help understand the correctness of \sysname, we highlight two important properties.

\parabf{The Safety Invariant}.
\emph{
Let $P$ be a predicate over the cluster state, e.g., Pod X is assigned to node Y, Pod Z is in Terminating, etc. If $P$ holds at a suffix of the sequence of controllers, then it eventually holds at all upstreams.
}

\parabf{The Liveness Assumption}.
\emph{
The narrow waist becomes totally connected infinitely often and sufficiently long for a round of end-to-end message passing.
}

\medskip
As described in \S\ref{sec:design:cache}, \sysname achieves the \emph{safety} invariant through the combination of soft and hard invalidation, which is guaranteed to propagate downstream changes upstream once two controllers are connected.
It ensures consistency between the upstream and downstream, upon which the upstream may reconcile and amend the cluster state in the face of failures.
The \emph{liveness} assumption ensures that the desired state can be eventually established through the entire narrow waist, allowing controllers to make progress.
It is reasonable because \sysname's message passing has millisecond-level latency (\S\ref{sec:eval:ablation}).
Furthermore, we enforce Pod lifecycle as a restriction on the transitions allowed during this process, guaranteeing end-to-end semantics.


\section{Implementation}
\label{sec:impl}

We implement \sysname on top of Kubernetes v1.32.0 with 3.8k lines of Go.
We modify $\sim$150 lines of code on average per controller in the narrow waist; the rest is for our common library.
We use the Knative FaaS platform for demonstration and retrofits its \emph{Autoscaler}.
As a proof of \sysname's compatibility, the \sysname-based Knative can directly work with the Prometheus~\cite{prometheus}, Kourier~\cite{kourier}, and Istio~\cite{istio} extensions for monitoring and networking.

\parabf{Exclusive ownership}.
\sysname has \emph{exclusive} ownership over the ephemeral state in the narrow waist.
Pods are protected because they remain hidden until the very end.
ReplicaSets and Deployments are not.
\sysname guards their {\lmtt{b}replicas} fields using Kubernetes admission control~\cite{k8s_webhook}.
External updates to the guarded fields will be rejected; non-essential fields such as annotations are unaffected.

\parabf{Pod discovery}.
Logically, one can obtain the routable FaaS endpoints by subscribing to the Pods that will be published from the narrow waist.
In practice, Kubernetes also offers the Service API~\cite{k8s_endpoints} that selects certain endpoints and abstracts them with a single static IP address.
Specifically, the \emph{Endpoints controller} monitors the Service selector and finds matching Pods. It then publishes the list of endpoints through the Endpoints API to the per-node \emph{Kube-proxies}, which handles address translation.
This process could also incur many API calls.
However, we note that the Endpoints are \emph{read-only} transformations of Pods, which do not require the design in \S\ref{sec:design}.
Therefore, we optimize the \emph{Endpoints controller} to directly stream the Endpoints to the \emph{Kube-proxies}.

\parabf{High-availability}.
\sysname is compatible with the high-availability (HA) setup of Kubernetes where controllers are replicated in primary-back mode~\cite{k8s_HA}.
This is because a controller becomes operational only if it wins the leader election, such that our assumption of a sequential structure still holds.
The new leader should run the handshake protocol upon takeover.


\section{Evaluation}
\label{sec:eval}

\begin{figure}[t]
\centering
\includegraphics[width=0.9\linewidth]{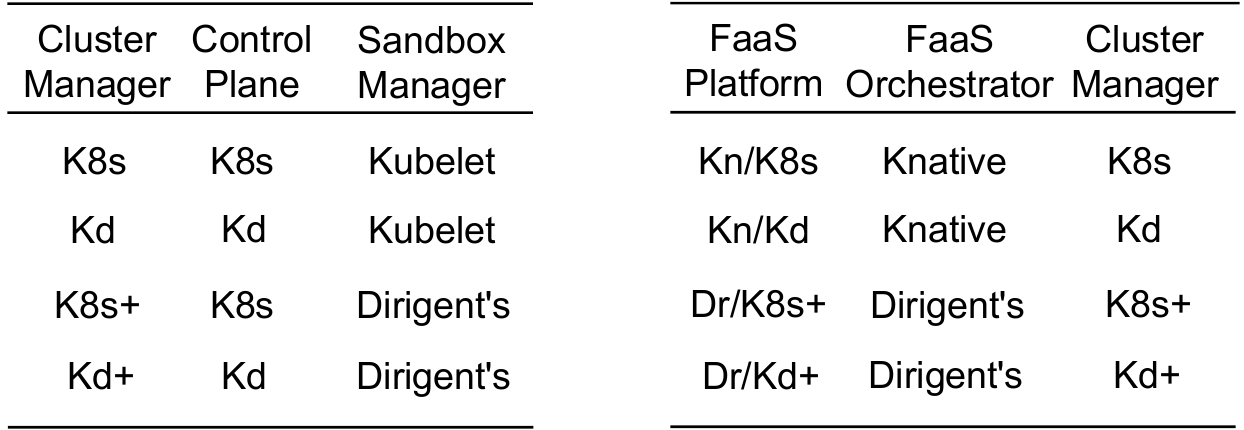}
\vspace{-0.1in}
\begin{subfigure}{0.49\linewidth}
\centering
\caption{Microbenchmarks}
\label{fig:eval:baselines:micro}
\end{subfigure}
\begin{subfigure}{0.49\linewidth}
\centering
\caption{End-to-End experiments}
\label{fig:eval:baselines:trace}
\end{subfigure}
\caption{Baselines in our evaluation.}
\vspace{-0.1in}
\label{fig:eval:baselines}
\end{figure}

\paraf{Methodology}.
We set up an 80-node cluster of xl170 instances on CloudLab~\cite{cloudlab}. Each node has ten Intel E5-2640v4 2.4 GHz CPU cores, 64 GB RAM, and 25 Gbps interconnect.

Our primary baselines are (1) the Kubernetes (K8s) v1.32.0 cluster manager, which is the codebase of \sysname (Kd), (2) Knative v1.15.0, a popular Kubernetes-based FaaS platform, and (3) Dirigent, the state-of-the-art clean-slate FaaS platform.
As per Figure~\ref{fig:bg:faas-arch}, a FaaS platform generally consists of the FaaS orchestrator and the base cluster manager; the latter consists of the control plane and the sandbox manager (e.g., \emph{Kubelet}) on worker nodes.
To understand the benefits of \sysname, we set up different combinations of FaaS orchestrators and sandbox managers as control variables as shown in Table~\ref{fig:eval:baselines}.
Note that it is possible to do so because \sysname focuses on message passing itself and do not depend on specific implementations.

For the cluster manager (Figure~\ref{fig:eval:baselines:micro}), Kd is the optimized K8s with direct message passing.
We also implement K8s+ and Kd+ by replacing the \emph{Kubelet} part of K8s and Kd with Dirigent's custom sandbox manager.
We run microbenchmarks on all variants to evaluate \sysname's primitives.

For the FaaS platform (Figure~\ref{fig:eval:baselines:trace}), we deploy Knative on K8s, denoted as Kn/K8s, and on Kd, denoted as Kn/Kd.
We also port Dirigent's orchestrator to K8s+ and Kd+, denoted as Dr/K8s+ and Dr/Kd+; Dr/K8s+ and Dr/Kd+ replace Dirigent with the respective control planes while retaining the orchestrator and sandbox manager.
We run end-to-end workloads on these baselines.
For fair comparison, we compare Kn/Kd with Kn/K8s, and Dr/Kd+ with Dr/K8s+ and Dirigent.

\subsection{Microbenchmarks}
\label{sec:eval:micro}

We scale out $N$ Pods for $K$ FaaS functions in an $M$-node cluster, and evaluate the scalability of the baselines over the three dimensions.
We use a strawman \emph{Autoscaler} that issues a one-shot scaling call per function, i.e., Deployment.
We measure the latency between the scaling call and the FaaS gateway receiving the complete list of ready Pods, as well as the time each controller spent during this process.

\begin{figure}[t]
\centering
\begin{subfigure}{\linewidth}
\centering
\includegraphics[width=\linewidth]{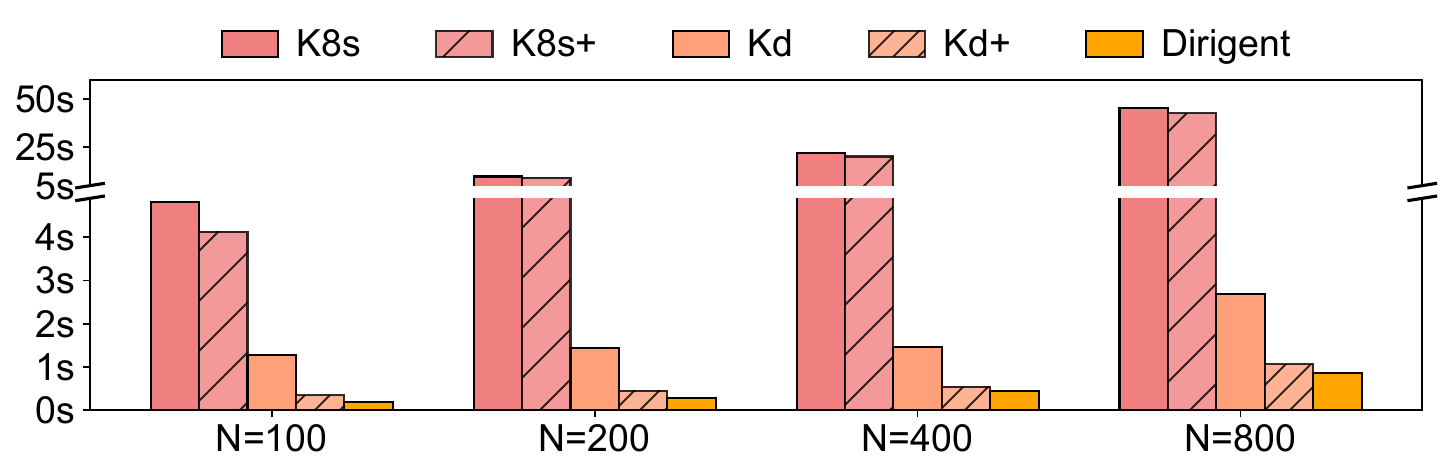}
\vspace{-0.25in}
\caption{End-to-end latency}
\label{fig:eval:scale-pods:e2e}
\end{subfigure}
\begin{subfigure}[b]{0.32\linewidth}
\centering
\includegraphics[width=\linewidth]{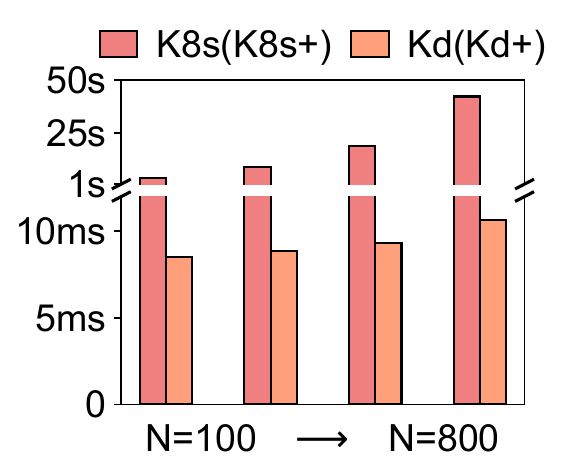}
\vspace{-0.25in}
\caption{ReplicaSet Ctrl.}
\label{fig:eval:scale-pods:rs}
\end{subfigure}
\hfill
\begin{subfigure}[b]{0.32\linewidth}
\centering
\includegraphics[width=\linewidth]{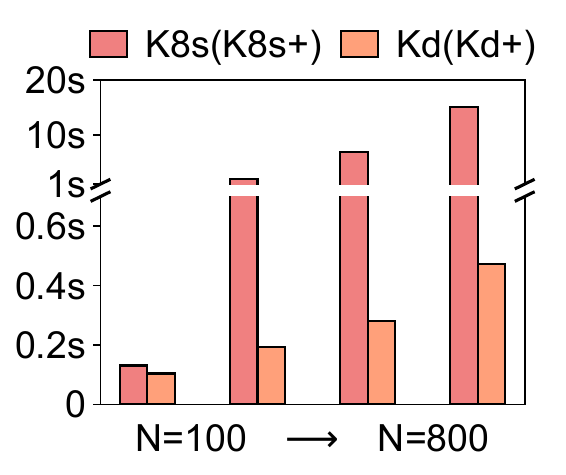}
\vspace{-0.25in}
\caption{Scheduler}
\label{fig:eval:scale-pods:sched}
\end{subfigure}
\hfill
\begin{subfigure}[b]{0.32\linewidth}
\centering
\includegraphics[width=\linewidth]{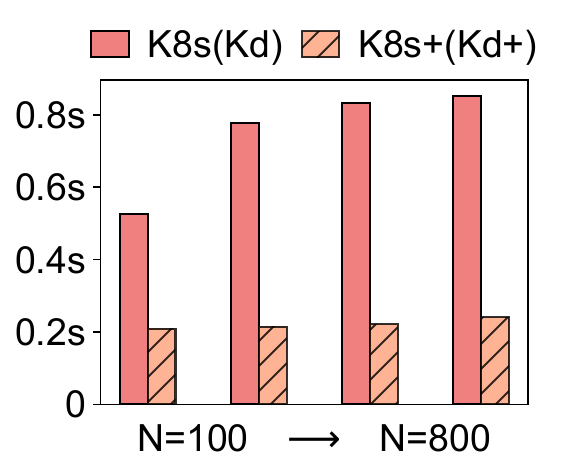}
\vspace{-0.25in}
\caption{Sandbox Mgr.}
\label{fig:eval:scale-pods:kubelet}
\end{subfigure}
\vspace{-0.1in}
\caption{Upscaling latency for varying \# of Pods.}
\vspace{-0.1in}
\label{fig:eval:scale-pods}
\end{figure}

\parabf{$N$-scalability (Pods)}.
We set $K$ to one and $M$ to 80, and vary $N$ from 100 to 800 Pods.
As shown in Figure~\ref{fig:eval:scale-pods:e2e}, \sysname significantly improves the scalability of the Kubernetes control plane.
Kd is 3.7--16.9$\times$ faster than K8s, and Kd+ 11.9--40.0$\times$ faster than K8s+.
K8s+ is marginally faster than K8s despite using Dirigents's optimized sandbox manager, because the bottleneck is in the control plane.
The benefit of a faster sandbox manager is only perceptible for Kd where the control plane is fast enough---Kd+ achieves the same sub-second latency as Dirigent, while being Kubernetes-compatible.

We further break down the end-to-end latency for the \emph{ReplicaSet controller}, the \emph{scheduler}, and the sandbox manager in Figure~\ref{fig:eval:scale-pods:rs}--\ref{fig:eval:scale-pods:kubelet}.
For the first two, K8s is equivalent to K8s+, and Kd is equivalent to Kd+.
For the sandbox manager, K8s is equivalent to Kd, and K8s+ is equivalent to Kd+.
Kd improves K8s by two orders of magnitude for the \emph{ReplicaSet controller}, and one for the \emph{scheduler}.
The improvements come from two factors.
First, Kd is free from the API rate-limiting in K8s.
Second, Kd's message passing itself has sub-millisecond latency, whereas it can take 10--35 milliseconds in K8s.
Consequently, \sysname approaches the raw performance of the internal control loops, adding up to 10\% overhead in our measurements.
The sandbox manager (Figure~\ref{fig:eval:scale-pods:kubelet}), in contrast, is relatively more scalable whether optimized or not.

We do not break down for the \emph{Autoscaler} or \emph{Deployment controller} because we set $K=1$ such that they both issue a single scaling call, with no scalability issue.
We study their performance next while we vary the number of functions $K$.

\begin{figure}[t]
\centering
\begin{subfigure}{\linewidth}
\centering
\includegraphics[width=\linewidth]{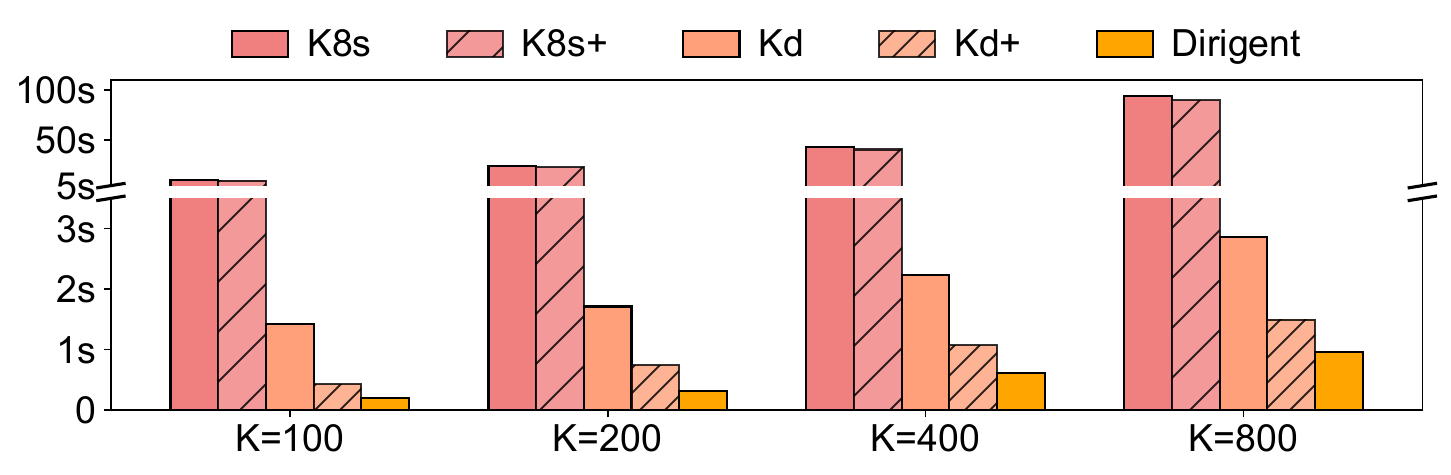}
\vspace{-0.25in}
\caption{End-to-end latency}
\label{fig:eval:scale-funcs:e2e}
\end{subfigure}
\begin{subfigure}[b]{0.32\linewidth}
\centering
\includegraphics[width=\linewidth]{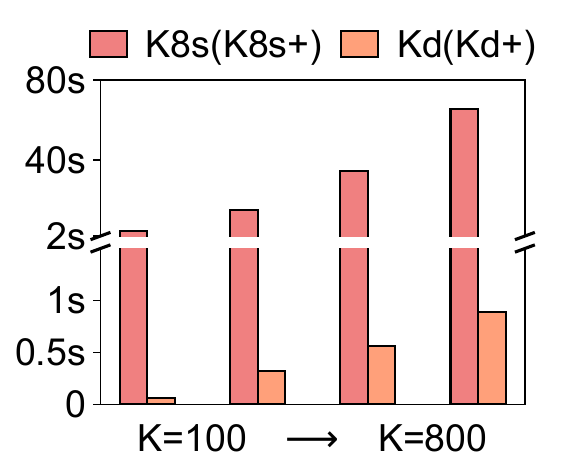}
\vspace{-0.25in}
\caption{Autoscaler}
\label{fig:eval:scale-funcs:as}
\end{subfigure}
\hfill
\begin{subfigure}[b]{0.32\linewidth}
\centering
\includegraphics[width=\linewidth]{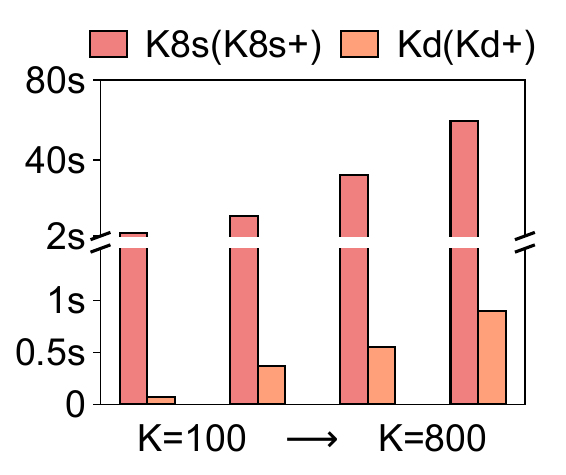}
\vspace{-0.25in}
\caption{Deployment Ctrl.}
\label{fig:eval:scale-funcs:dp}
\end{subfigure}
\hfill
\begin{subfigure}[b]{0.32\linewidth}
\centering
\includegraphics[width=\linewidth]{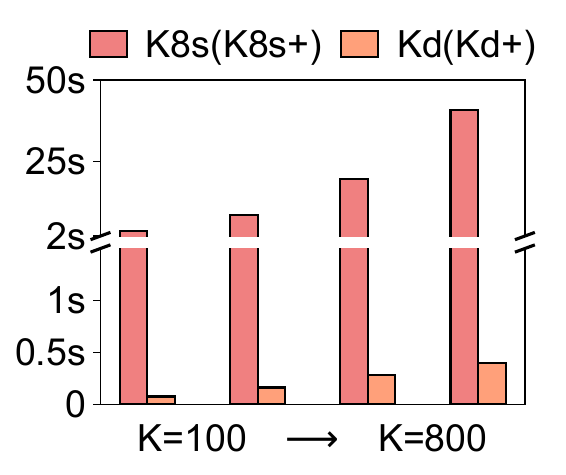}
\vspace{-0.25in}
\caption{ReplicaSet Ctrl.}
\label{fig:eval:scale-funcs:rs}
\end{subfigure}
\vspace{-0.1in}
\caption{Upscaling latency for varying \# of functions.}
\vspace{-0.05in}
\label{fig:eval:scale-funcs}
\end{figure}

\parabf{$K$-scalability (Functions)}.
We set $M$ to 80, and vary $K$ from 100 to 800 functions with one Pod per function, i.e., $N=K$.
We omit the breakdown for the \emph{scheduler} and the sandbox manager because there is no difference from previous case in their perspective.
As shown in Figure~\ref{fig:eval:scale-funcs:e2e}, Kd is 7.4--32.8$\times$ faster than K8s, and Kd+ 22.7--59.8$\times$ faster than K8s+.
K8s (K8s+) experiences higher latency than that in Figure~\ref{fig:eval:scale-pods}, because it has to scale up the same number of Pods while doing it on a per-function basis.
This increases the message passing overhead for the \emph{Autoscaler}
and the \emph{Deployment controller} (Figure~\ref{fig:eval:scale-funcs:as} and \ref{fig:eval:scale-funcs:dp}).
\sysname improves the former by 39.5--70.3$\times$ and the latter by 31.2--50.1$\times$.
In terms of the \emph{ReplicaSet controller} (Figure~\ref{fig:eval:scale-funcs:rs}), K8s's performance is similar to that in Figure~\ref{fig:eval:scale-pods}; Kd's latency is increased by $60-350$ milliseconds because per-function scaling results in less batching.
However, the impact on the end-to-end latency is marginal due to inter-controller pipelining.

\begin{figure}[t]
\centering
\includegraphics[width=0.8\linewidth]{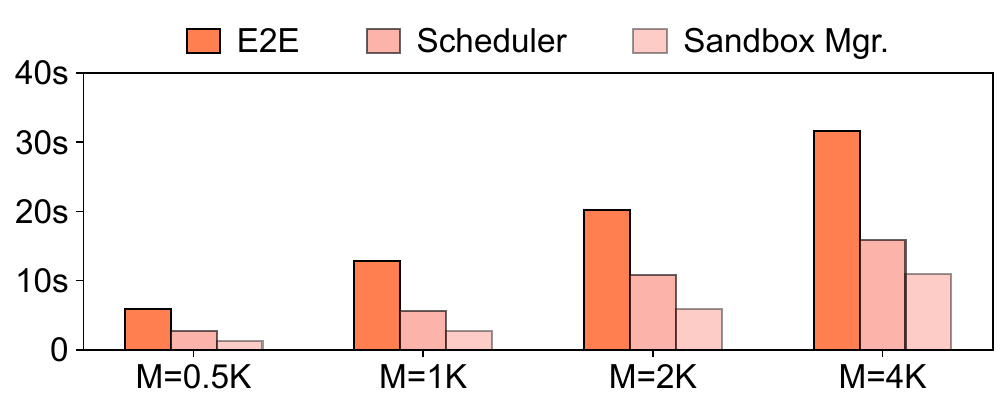}
\vspace{-0.15in}
\caption{Upscaling latency for varying \# of nodes.}
\vspace{-0.1in}
\label{fig:eval:scale-nodes}
\end{figure}

\parabf{$M$-scalability (nodes)}.
Finally, we evalute \sysname's scalability in large clusters. We vary the number of nodes $M$ from 500 to 4000 and scale up five Pods per node.
Because we do not have a large enough real cluster, we set up fake sandbox managers per node that simulates container startup but \emph{does} expose the Pods through Kubernetes API.
Figure~\ref{fig:eval:scale-nodes} shows that Kd can scale up 20K Pods in 30 seconds.
We see a substantial increase of latency in the \emph{Scheduler} because its algorithm has to consider more nodes for the placement of Pods.
\sysname's message passing adds up to 6\% overhead to this process.
The sandbox manager also experiences higher latency due to the $\sim$20K concurrent API calls as the Pods are exposed.
Note that each sandbox manager still follows its configured API rate limits; the high load on the API Server is a problem inherent to large Kubernetes clusters~\cite{apiserver_latency}, which limits the maximum cluster size to 5000 nodes~\cite{k8s_cluster_limit}.

\parabf{Downscaling}.
We also evaluate the downscaling performance of \sysname under the same setups.
Because the number of API calls or messages required is approximately the same as upscaling, we observe similar characteristics in performance.
Specifically, for $K$-scalability, Kd is 6.9--30.3$\times$ faster than K8s, and Kd+ 16.8--45.2$\times$ faster than K8s+.

\begin{figure}[t]
\centering
\includegraphics[width=\linewidth]{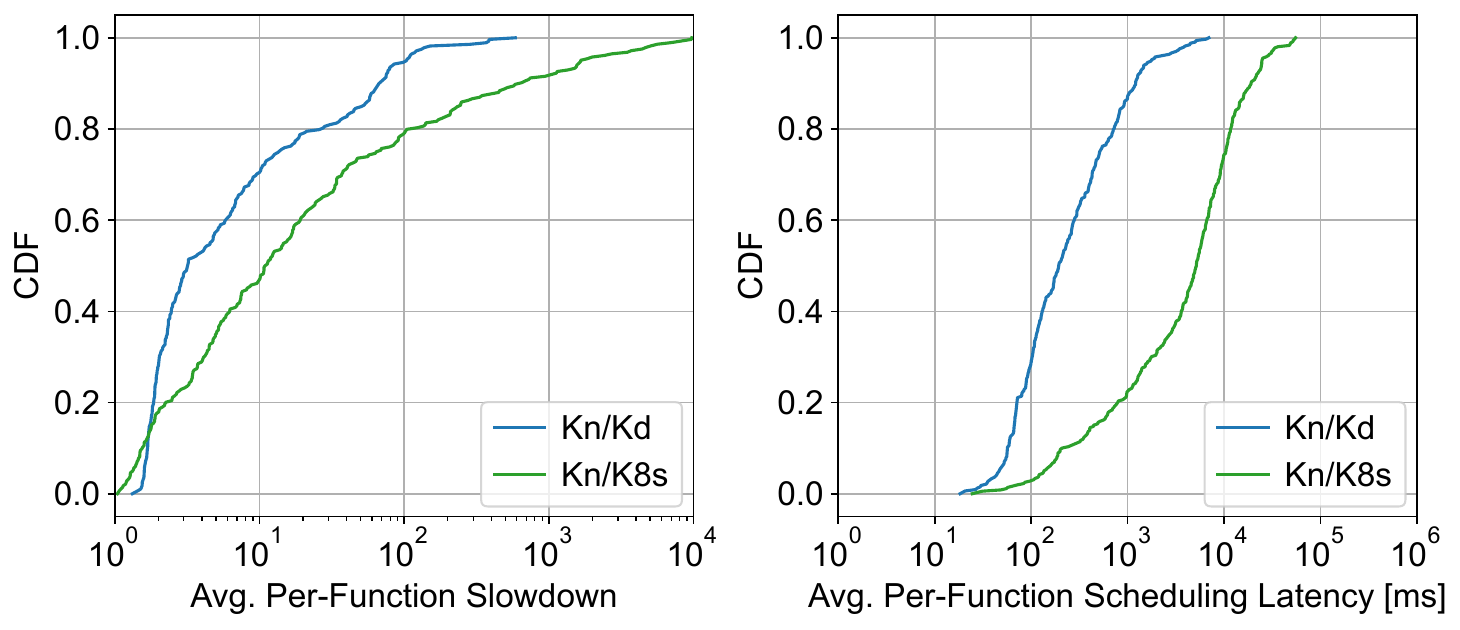}
\vspace{-0.2in}
\caption{End-to-end performance on Knative-variants.}
\label{fig:eval:trace:knative}
\end{figure}

\begin{figure}[t]
\centering
\includegraphics[width=\linewidth]{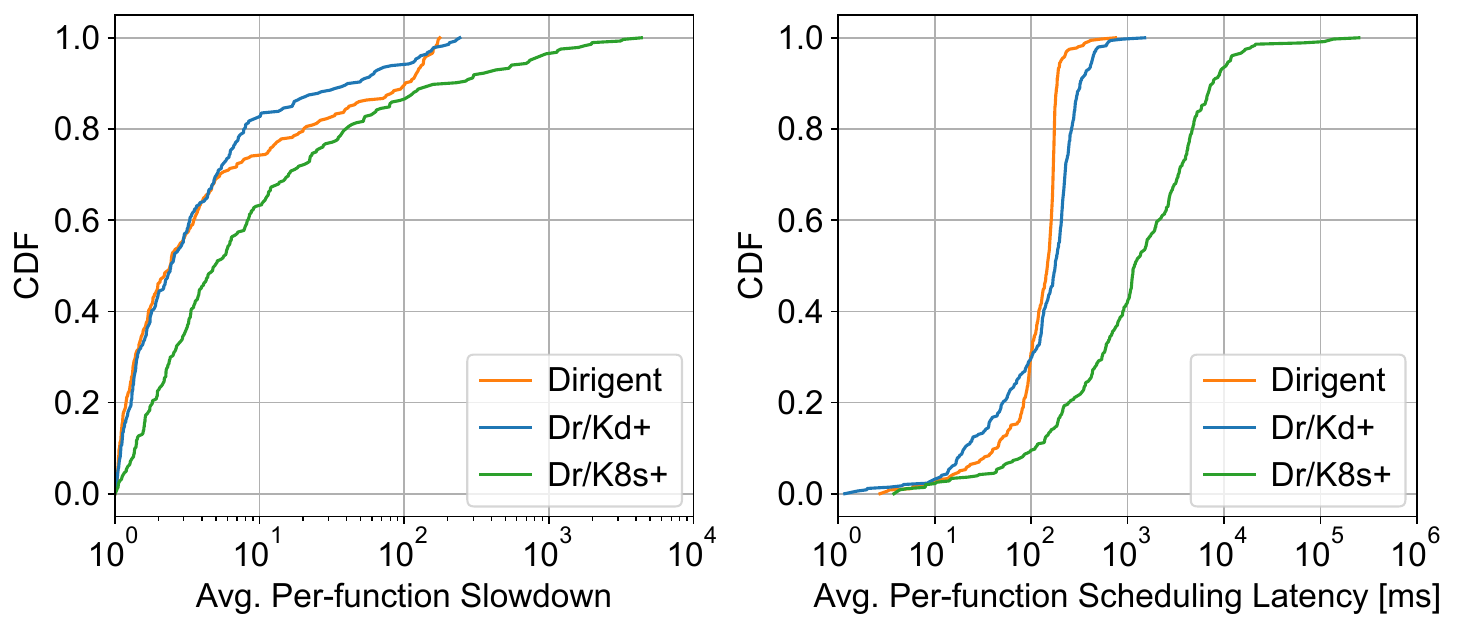}
\vspace{-0.2in}
\caption{End-to-end performance on Dirigent-variants.}
\label{fig:eval:trace:dirigent}
\end{figure}

\subsection{End-to-End FaaS Workload}
\label{sec:eval:e2e}

We run the same FaaS workload as Dirigent (available in its artifact~\cite{dirigent_artifact}).
Specifically, it is a 30-minute clip of the Microsoft Azure Functions trace~\cite{shahrad2020serverless} with 500 functions and 168K invocations.
The durations of invocations are sampled based on the percentiles given in the trace; to handle an invocation, a function instance busy loops with the SQRTSD x86 instruction based on the requested duration.

In line with Dirigent, we consider two performance metrics: the average request slowdown and the average request scheduling latency.
The former is end-to-end request latency divided by the requested execution time, and the latter is the time from the invocation arrival to the beginning of its processing by some function instance.
Because the execution times and invocation frequencies of different functions in the trace can vary by orders of magnitude, we group the metrics by function and plot the overall CDF.

Figure~\ref{fig:eval:trace:knative} and \ref{fig:eval:trace:dirigent} shows the results for baselines in Table~\ref{fig:eval:baselines:trace}.
For the Knative-variants, Kn/Kd improves the median (p99) slowdown by 3.5$\times$ (19.4$\times$) and the median (p99) scheduling latency by 26.7$\times$ (10.3$\times$) compared to Kn/K8s.
For the Dirigent-variants, Dr/Kd+ improves the median (p99) slowdown by 2.0$\times$ (10.4$\times$) and the median (p99) scheduling latency by 6.6$\times$ (134$\times$) compared to Dr/K8s+.
Despite using the Kubernetes code base and retaining compatibility, Dr/Kd+ achieves similar performance as Dirigent.

We find that the long tails of Kn/K8s and Dr/K8s+ come from periodical bursts of \emph{cold} functions, i.e., they are infrequent as individuals, but tend to arrive simultaneously~\cite{dirigent,shahrad2020serverless}.
This is no longer a problem for Kn/Kd, Dr/Kd+, or Dirigent, because the control plane can more efficiently absorb the bursts; instead, the tails of these baselines are caused by functions with very short durations.
We also observe a 67\% reduction in the number of cold starts when replacing K8s with Kd.
This is due to the autoscaling policy of Knative and Dirigent that computes the desired replicas count based on the number of inflight requests.
Faster upscaling can effectively reduce the queuing effect and prevent the \emph{Autoscaler} from desperately scaling up even more replicas.

\begin{figure}[t]
\centering
\includegraphics[width=0.7\linewidth]{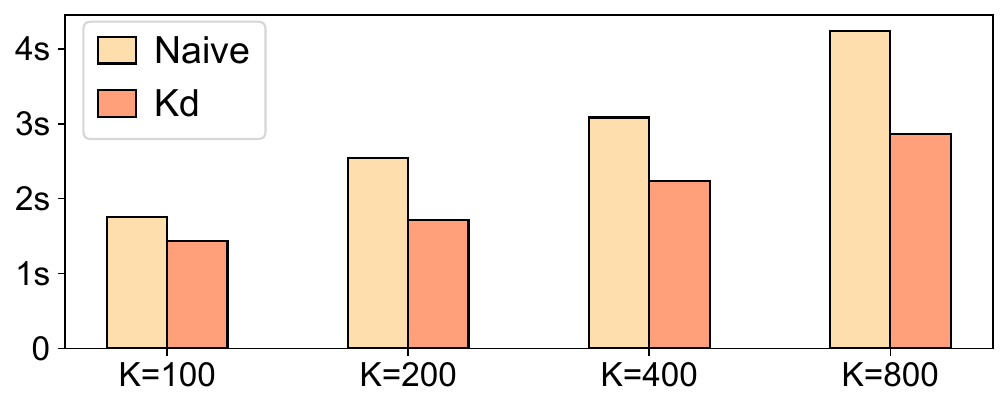}
\vspace{-0.1in}
\caption{Benefits of on-demand materialization.}
\label{fig:eval:abl:on-demand}
\end{figure}

\begin{figure}[t]
\centering
\begin{subfigure}[b]{0.32\linewidth}
\centering
\includegraphics[width=\linewidth]{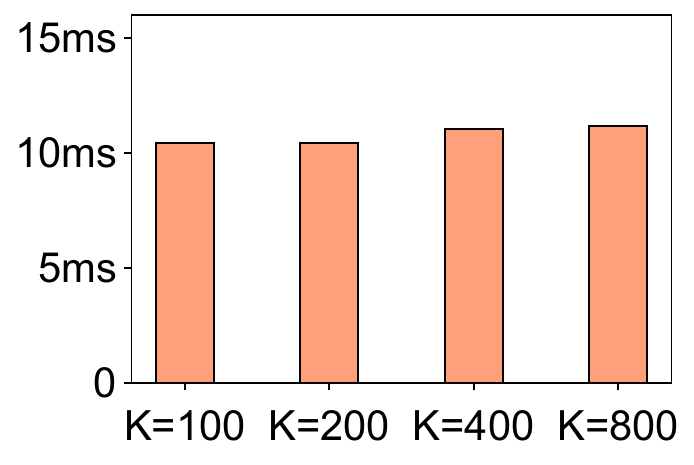}
\vspace{-0.2in}
\caption{Autoscaler}
\label{fig:eval:abl:handshake:as}
\end{subfigure}
\hfill
\begin{subfigure}[b]{0.32\linewidth}
\centering
\includegraphics[width=\linewidth]{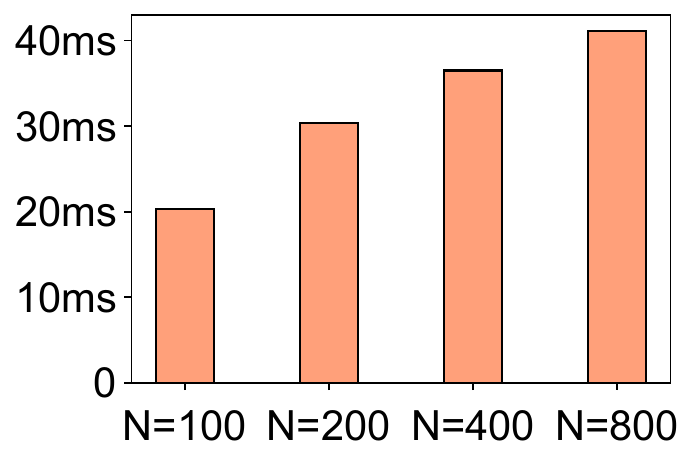}
\vspace{-0.2in}
\caption{ReplicaSet Ctrl.}
\label{fig:eval:abl:handshake:rs}
\end{subfigure}
\hfill
\begin{subfigure}[b]{0.32\linewidth}
\centering
\includegraphics[width=\linewidth]{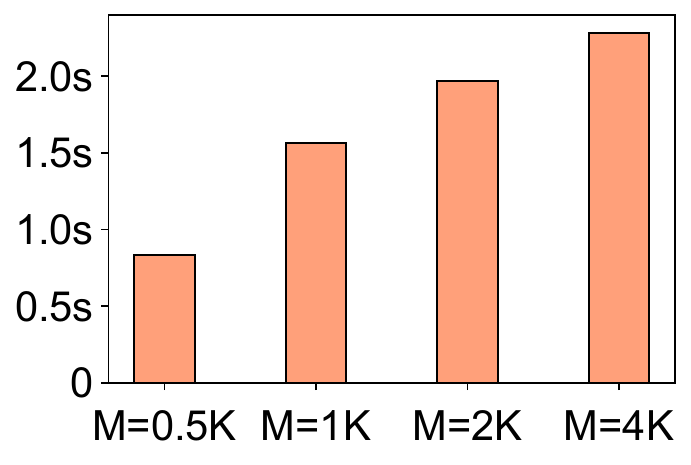}
\vspace{-0.2in}
\caption{Scheduler}
\label{fig:eval:abl:handshake:sched}
\end{subfigure}
\vspace{-0.1in}
\caption{Failure handling with hard invalidation.}
\vspace{-0.1in}
\label{fig:eval:abl:handshake}
\end{figure}

\subsection{Ablation Study}
\label{sec:eval:ablation}

\paraf{Dynamic materialization}.
To understand the necessity of on-demand materialization, we compare it with naive direct message passing that sends full API objects; it avoids the overhead of persistence but not serialization and deserialization.
We use the $K$-scalability setup in \S\ref{sec:eval:micro}.
Figure~\ref{fig:eval:abl:on-demand} shows the naive approach incurs 20--35\% higher latency than \sysname, which justifies our design choice.

\parabf{Hard invalidation}.
We measure the overhead of hard invalidation, i.e., the handshake protocol, by resetting controller cache as if in crash-restarts and thus forcing handshakes.
We use the $K$-scalability setup \S\ref{sec:eval:micro} to populate the cache of the \emph{Autoscaler} and \emph{Deployment controller}, and use the $N-$, and $M$-scalability setups for the \emph{ReplicaSet controller}, and \emph{Scheduler} respectively.
Figure~\ref{fig:eval:abl:handshake} shows the result.
The \emph{Autoscaler} and \emph{Deployment controller} (not shown) have similar and neglectable overhead; this is because they are level-triggered and idempotent, so we asynchronously perform state forwarding in subsequent scaling calls.
The \emph{ReplicaSet controller} has sub-linear overhead because Pods can be exchanged in batches.
The \emph{Scheduler} also has sub-linear overhead because it handshakes with the \emph{Kubelets} in parallel.

\parabf{Soft invalidation}.
We measure the overhead of soft invalidations and its impact on \emph{synchronous} termination, e.g., preemption, which needs to wait for downstream signals (\S\ref{sec:design:lifecycle}).
Specifically, one hop of soft invalidation takes 0.5--1.2 milliseconds, similar to forward message passing.
The end-to-end preemption latency, including two hops and the processing at the \emph{Kubelet}, is 6.2--13.4 milliseconds, compared to the 10--35 milliseconds latency of a standard API call.


\section{Discussion}
\label{sec:discussion}

\paraf{Deploying \sysname}.
\sysname cannot be developed as a bolt-on extension to Kubernetes because the \emph{Scheduler} and the \emph{Kubelet} must be aware of the ephemeral Pods to correctly allocate and share resources.
However, it does support a rollout deployment by gradually replacing the controllers with \sysname's, in the \emph{reverse} order of the narrow waist.
Because \sysname is fully compatible with the Kubernetes APIs, the downtime can be kept to a minimum.
We also expect long-term compatibility with future Kubernetes releases, because the narrow waist has been stable through many years of development.

\parabf{Horizontal scaling}.
\sysname vertically scales the capability of a single cluster. An orthogonal approach is to horizontally scale out multiple sub-clusters to amortize the overhead.
However, based on Figure~\ref{fig:bg:gap}, handling the cold starts of the Azure trace requires $\sim$50 sub-clusters, leading to management burdens, load imbalance, and resource fragmentation~\cite{dirigent}.
While \sysname can also scale horizontally, the fact that it is 30$\times$ faster than Kubernetes means it needs much fewer sub-clusters for the same throughput.
Another approach is to replicate and shard the controllers in the narrow waist (\S\ref{sec:bg:bottleneck}). However, doing so requires non-trivial code changes, even more configuring and tuning, and risks load imbalance and conflicts across shards, which contradicts our design goal of minimum intrusion and maximum reuse.

\parabf{Stability}.
While \sysname increases the upscaling throughput, the \emph{Kubelets} still need to call the Kubernetes API. However, this will not compromise the stability of the API Server because the \emph{Kubelets} still follow the API rate limits.

\parabf{Availability}.
Although \sysname requires direct connections between controllers, it typically offers the same availability as the decoupled approach of Kubernetes.
In both systems, a controller can only make substantial progress when its desired state can be processed by its downstream.
Also note that disconnections between controller will not affect the progress of non-\sysname managed objects, because we enforce a clean separation of ownership.

\parabf{Observability}.
The ephemeral Pods in \sysname are invisible to external controllers util they exit the narrow waist.
However, our design of the flexible message format (\S\ref{sec:overview:dynmat}) and handshake protocol (\S\ref{sec:design:cache}) allows the narrow waist to be extensible.
Cluster operators can insert monitoring modules into the narrow waist to observe individual events of Pod creation and scheduling.

\parabf{Broader applicability}.
\sysname is built on top of Kubernetes but not limited to it.
Many cluster managers have also adopted the state-centric approach like Kubernetes~\cite{verma2015large,tang2020twine,schwarzkopf2013omega}, so they can also benefit from direct message passing.
\sysname can also extends to custom controllers because they uniformly follow the state-centric workflow (Figure~\ref{fig:overview:arch}), and \sysname is minimally intrusive to their internal logic.


\section{Related Work}
\label{sec:related}

\paraf{Cluster managers}
are responsible for governing the cloud infrastructure~\cite{openstack, eric2014apollo, delimitrou2015tarcil, gog2016firmament, hindman2011mesos, isard2009quincy}.
Modularity and extensibility are important design principles.
Kubernetes~\cite{k8s}, Borg~\cite{verma2015large}, Twine~\cite{tang2020twine}, Omega~\cite{schwarzkopf2013omega}, and vSphere~\cite{vsphere} adopt state-centric APIs to break down the control plane into loosely coupled controllers.
However, they are traditionally designed for long-running workloads where orchestration can be amortized over time~\cite{hindman2011mesos, eric2014apollo}, contradicting the bursty and short-lived nature of FaaS~\cite{liu2023gap}.
\sysname presents a practical design to bridge this gap.

\parabf{FaaS orchestration}
is an active research area~\cite{zhang2021faster,wang2018peeking,jegan2023guarding,chaiForkRoadReflections2025}.
Many works present new autoscaling, load-balancing, and placement algorithms and realize them on top of existing cluster managers~\cite{abdi2023palette,bilal2023great,liu2024jiagu,stojkovic2024ecofaas,stojkovic2023mxfaas,kuchler2023function,alzayat2023groundhog}.
Dirigent~\cite{dirigent} and others~\cite{fuerst2023iluvatar,singhvi2021atoll,chen2024yuanrong,szekely2024unifying,mittal2021mu} identify the mismatch between FaaS and legacy managers and propose novel clean-slate architectures.
\sysname is inspired by their observations and boils down the problem to message passing.
It retains existing architecture and APIs for compatibility.

\parabf{State machine replication} is a common approach in building fault-tolerant distributed services~\cite{aguilera2020microsecond,cui2015paxos,enes2021efficient,pan2021rabia,balakrishnan2020virtual,balakrishnan2021log,ganesan2021exploiting,lockerman2018fuzzylog,luo2024lazylog}. It ensures that commands are executed in a consistent order across replicas. \sysname is particularly inspired by the Chain Replication protocol~\cite{vanrenesse2004chain} that assumes a sequential structure like the narrow waist.
However, \sysname is faced with a novel and more complicated problem where the state machines are heterogeneous and can progressively modify their state.

\parabf{Speculative execution}
is a key idea behind \sysname's fast upscaling, where controllers opportunistically drive the cluster to its desired state instead of committing step by step.
This technique has been used in a wide context ranging from distributed systems~\cite{li2022speculative,ganesan2021exploiting,luo2024lazylog}, FaaS serving~\cite{stojkovic2023specfaas}, to large language model inference~\cite{miao2024specinfer,tan2025pipellm}.
The idea is to replace expensive operations with cheap ones that do not \emph{guarantee} but \emph{very likely} provide the desired property, and implement safeguard mechanisms to ensure correctness.


\section{Conclusion}
\label{sec:conclusion}

\sysname retrofits Kubernetes to handle bursty FaaS workloads.
The common narrow waist structure across FaaS platforms allows us to achieve both efficiency and compatibility.
We employ lightweight direct message passing between controllers to bypass the API Server bottleneck.
In spite of the asynchrony of controllers, ephemerality of their state, and the non-idempotence of their operations, we implement a hierarchical write-back cache through the narrow waist that provides a consistent view of cluster state.
In addition, we specifically handle termination to comply with the convention on instance lifecycle.
\sysname has similar performance as the state-of-the-art system Dirigent.

\bibliographystyle{plain}
\bibliography{urls,references}

\clearpage
\appendix
\pagestyle{plain}
\onecolumn

\section*{TLA+ Specification of \sysname}

\begin{tlatex}
\@x{}\moduleLeftDash\@xx{ {\MODULE} KdInterface}\moduleRightDash\@xx{}%
\@x{ {\EXTENDS} Naturals ,\, Sequences ,\, FiniteSets}%
\@pvspace{8.0pt}%
\@x{ {\CONSTANTS}}%
\@x{\@s{16.4} ScalingCmds ,\,}%
\@x{\@s{16.4} NumNodes}%
\@pvspace{8.0pt}%
\@x{ {\VARIABLES}}%
\@x{\@s{16.4} CmdIndex ,\,}%
\@x{\@s{16.4} APIPods}%
\@pvspace{8.0pt}%
\@x{ {\VARIABLES}}%
\@x{\@s{16.4} AsRsConn ,\,}%
\@x{\@s{16.4} RsSchedConn ,\,}%
\@x{\@s{16.4} SchedKletConn}%
\@pvspace{8.0pt}%
\@x{ ConstantsTypeOK \.{\defeq}}%
\@x{\@s{16.4} \.{\land} NumNodes \.{\in} Nat \.{\,\backslash\,} \{\@w{0} \}}%
 \@x{\@s{16.4} \.{\land} ScalingCmds \.{\in} Seq ( Nat \.{\,\backslash\,}
 \{\@w{0} \} )}%
\@x{\@s{16.4} \.{\land} Len ( ScalingCmds ) \.{>} 0}%
 \@x{\@s{16.4} \.{\land} \A\, i \.{\in} 1 \.{\dotdot} Len ( ScalingCmds )
 \.{-} 1 \.{:} ScalingCmds [ i ] \.{<} ScalingCmds [ i \.{+} 1 ]}%
\@pvspace{8.0pt}%
\@x{ APITypeOK \.{\defeq}}%
\@x{\@s{16.4} \.{\land} CmdIndex \.{\in} 1 \.{\dotdot} Len ( ScalingCmds )}%
 \@x{\@s{16.4} \.{\land} APIPods \.{\in} {\SUBSET} [ pod \.{:} Nat
 \.{\,\backslash\,} \{\@w{0} \} ,\, node \.{:} 1 \.{\dotdot} NumNodes ]}%
 \@x{\@s{16.4} \.{\land} \A\, p1 ,\, p2 \.{\in} APIPods \.{:} p1 . pod \.{=}
 p2 . pod \.{\implies} p1 . node \.{=} p2 . node}%
\@pvspace{8.0pt}%
\@x{ ConnTypeOK \.{\defeq}}%
 \@x{\@s{16.4} \.{\land} AsRsConn \.{\in} [ inflight \.{:} Seq ( Nat
 \.{\,\backslash\,} \{\@w{0} \} ) ,\, connected \.{:} {\BOOLEAN} ]}%
 \@x{\@s{16.4} \.{\land} RsSchedConn \.{\in} [ inflight \.{:} Seq ( [ batchId
 \.{:} Nat ,\, batchSize \.{:} Nat \.{\,\backslash\,} \{\@w{0} \} ] ) ,\,
 connected \.{:} {\BOOLEAN} ]}%
 \@x{\@s{16.4} \.{\land} SchedKletConn \.{\in} [ inflight \.{:} Seq ( [ pod
 \.{:} Nat \.{\,\backslash\,} \{\@w{0} \} ,\, node \.{:} 1 \.{\dotdot}
 NumNodes ] ) ,\, connected \.{:} {\BOOLEAN} ]}%
\@pvspace{8.0pt}%
\@x{ InterfaceOK \.{\defeq}}%
\@x{\@s{16.4} \.{\land} ConstantsTypeOK}%
\@x{\@s{16.4} \.{\land} APITypeOK}%
\@x{\@s{16.4} \.{\land} ConnTypeOK}%
\@pvspace{8.0pt}%
\@x{}\midbar\@xx{}%
\@pvspace{8.0pt}%
\@x{ APIUnchanged \.{\defeq}}%
\@x{\@s{16.4} {\UNCHANGED} {\langle} CmdIndex ,\, APIPods {\rangle}}%
\@pvspace{8.0pt}%
\@x{ AsRsDisconnected \.{\defeq}}%
\@x{\@s{16.4} {\IF} AsRsConn . connected \.{=} {\TRUE} \.{\THEN}}%
 \@x{\@s{32.65} AsRsConn \.{'} \.{=} [ inflight \.{:} {\langle} {\rangle} ,\,
 connected \.{:} {\FALSE} ]}%
\@x{\@s{16.4} \.{\ELSE}}%
\@x{\@s{32.8} {\UNCHANGED} AsRsConn}%
\@pvspace{8.0pt}%
\@x{ RsSchedDisconnected \.{\defeq}}%
\@x{\@s{16.4} {\IF} RsSchedConn . connected \.{=} {\TRUE} \.{\THEN}}%
 \@x{\@s{32.65} RsSchedConn \.{'} \.{=} [ inflight \.{:} {\langle} {\rangle}
 ,\, connected \.{:} {\FALSE} ]}%
\@x{\@s{16.4} \.{\ELSE}}%
\@x{\@s{32.8} {\UNCHANGED} RsSchedConn}%
\@pvspace{8.0pt}%
\@x{ SchedKletDisconnected \.{\defeq}}%
\@x{\@s{16.4} {\IF} SchedKletConn . connected \.{=} {\TRUE} \.{\THEN}}%
 \@x{\@s{32.65} SchedKletConn \.{'} \.{=} [ inflight \.{:} {\langle} {\rangle}
 ,\, connected \.{:} {\FALSE} ]}%
\@x{\@s{16.4} \.{\ELSE}}%
\@x{\@s{32.8} {\UNCHANGED} SchedKletConn}%
\@pvspace{8.0pt}%
\@x{}\midbar\@xx{}%
\@pvspace{8.0pt}%
\@x{ DoAPIInit \.{\defeq}}%
\@x{\@s{16.4} \.{\land} CmdIndex \.{=} 1}%
\@x{\@s{16.4} \.{\land} APIPods \.{=} \{ \}}%
\@pvspace{8.0pt}%
\@x{ DoConnInit \.{\defeq}}%
 \@x{\@s{16.4} \.{\land} AsRsConn \.{=} [ inflight \.{:} {\langle} {\rangle}
 ,\, connected \.{:} {\TRUE} ]}%
 \@x{\@s{16.4} \.{\land} RsSchedConn \.{=} [ inflight \.{:} {\langle}
 {\rangle} ,\, connected \.{:} {\TRUE} ]}%
 \@x{\@s{16.4} \.{\land} SchedKletConn \.{=} [ inflight \.{:} {\langle}
 {\rangle} ,\, connected \.{:} {\TRUE} ]}%
\@pvspace{8.0pt}%
\@x{ DoInterfaceInit \.{\defeq}}%
\@x{\@s{16.4} \.{\land} DoAPIInit}%
\@x{\@s{16.4} \.{\land} DoConnInit}%
\@pvspace{8.0pt}%
\@x{}\midbar\@xx{}%
\@pvspace{8.0pt}%
\@x{ {\RECURSIVE} SetToSeq ( \_ )}%
\@x{ SetToSeq ( S ) \.{\defeq}}%
\@x{\@s{8.2} {\IF} S \.{=} \{ \} \.{\THEN} {\langle} {\rangle}}%
 \@x{\@s{8.2} \.{\ELSE} \.{\LET} x \.{\defeq} {\CHOOSE} x \.{\in} S \.{:}
 {\TRUE}}%
 \@x{\@s{39.51} \.{\IN} {\langle} x {\rangle} \.{\circ} SetToSeq ( S
 \.{\,\backslash\,} \{ x \} )}%
\@pvspace{8.0pt}%
\@x{}\bottombar\@xx{}%
\@pvspace{8.0pt}%
\@x{}\moduleLeftDash\@xx{ {\MODULE} Autoscaler}\moduleRightDash\@xx{}%
\@x{ {\EXTENDS} KdInterface}%
\@pvspace{8.0pt}%
\@x{ {\VARIABLES}}%
\@x{\@s{16.4} LastDesiredReplicas}%
\@pvspace{8.0pt}%
\@x{ AsTypeOK \.{\defeq}}%
\@x{\@s{16.4} \.{\land} LastDesiredReplicas \.{\in} Nat}%
 \@x{\@s{16.4} \.{\land} LastDesiredReplicas \.{\leq} ScalingCmds [ CmdIndex
 ]}%
\@pvspace{8.0pt}%
\@x{ AsOK \.{\defeq} AsTypeOK}%
\@pvspace{8.0pt}%
 \@x{ AsUnchanged \.{\defeq} {\UNCHANGED} {\langle} LastDesiredReplicas
 {\rangle}}%
\@pvspace{8.0pt}%
\@x{}\midbar\@xx{}%
\@pvspace{8.0pt}%
\@x{ DoAsInit \.{\defeq}}%
\@x{\@s{16.4} LastDesiredReplicas \.{=} 0}%
\@pvspace{8.0pt}%
\@x{ DoAsCrash \.{\defeq}}%
\@x{\@s{16.4} \.{\land} LastDesiredReplicas \.{'} \.{=} 0}%
\@x{\@s{16.4} \.{\land} APIUnchanged}%
\@x{\@s{16.4} \.{\land} AsRsDisconnected}%
 \@x{\@s{16.4} \.{\land} {\UNCHANGED} {\langle} RsSchedConn ,\, SchedKletConn
 {\rangle}}%
\@pvspace{8.0pt}%
\@x{ DoAsControlLoop \.{\defeq}}%
\@x{\@s{16.4} \.{\LET} trySend ( scale ) \.{\defeq}}%
\@x{\@s{53.19} {\IF} AsRsConn . connected \.{=} {\TRUE} \.{\THEN}}%
 \@x{\@s{69.45} \.{\land} AsRsConn \.{'} \.{=} [ AsRsConn {\EXCEPT} {\bang} .
 inflight \.{=} @ \.{\circ} {\langle} scale {\rangle} ]}%
\@x{\@s{69.45} \.{\land} LastDesiredReplicas \.{'} \.{=} scale}%
\@x{\@s{53.19} \.{\ELSE}}%
\@x{\@s{69.59} \.{\land} {\UNCHANGED} LastDesiredReplicas}%
\@x{\@s{69.59} \.{\land} {\UNCHANGED} AsRsConn}%
\@x{\@s{36.79} tryLoopOnce \.{\defeq}}%
 \@x{\@s{53.19} {\IF} CmdIndex \.{\in} 1 \.{\dotdot} Len ( ScalingCmds )
 \.{\land} LastDesiredReplicas \.{\neq} ScalingCmds [ CmdIndex ] \.{\THEN}}%
\@x{\@s{69.45} trySend ( ScalingCmds [ CmdIndex ] )}%
\@x{\@s{53.19} \.{\ELSE}}%
\@x{\@s{69.59} \.{\land} {\UNCHANGED} LastDesiredReplicas}%
\@x{\@s{69.59} \.{\land} {\UNCHANGED} AsRsConn}%
\@x{\@s{16.4} \.{\IN}}%
\@x{\@s{16.4} \.{\land} tryLoopOnce}%
\@x{\@s{16.4} \.{\land} APIUnchanged}%
 \@x{\@s{16.4} \.{\land} {\UNCHANGED} {\langle} RsSchedConn ,\, SchedKletConn
 {\rangle}}%
\@pvspace{8.0pt}%
\@x{}\bottombar\@xx{}%
\@pvspace{8.0pt}%
\@x{}\moduleLeftDash\@xx{ {\MODULE} ReplicaSet}\moduleRightDash\@xx{}%
\@x{ {\EXTENDS} KdInterface}%
\@pvspace{8.0pt}%
\@x{ {\VARIABLES}}%
\@x{\@s{16.4} DesiredReplicas ,\,}%
\@x{\@s{16.4} CreatedPods ,\,}%
\@x{\@s{16.4} MaxPodId}%
\@pvspace{8.0pt}%
\@x{ RsTypeOK \.{\defeq}}%
\@x{\@s{16.4} \.{\land} MaxPodId \.{\in} Nat}%
\@x{\@s{16.4} \.{\land} DesiredReplicas \.{\in} Nat}%
\@x{\@s{16.4} \.{\land} DesiredReplicas \.{\leq} ScalingCmds [ CmdIndex ]}%
\@x{\@s{16.4} \.{\land} CreatedPods \.{\in} {\SUBSET} Nat}%
 \@x{\@s{16.4} \.{\land} \A\, p \.{\in} CreatedPods \.{:} p \.{\in} 1
 \.{\dotdot} MaxPodId}%
\@pvspace{8.0pt}%
\@x{ RsPodUnique \.{\defeq} {\TRUE}}%
\@pvspace{8.0pt}%
\@x{ RsPodComplete \.{\defeq}}%
\@x{\@s{16.4} \A\, p \.{\in} APIPods \.{:} p \.{\in} CreatedPods}%
\@pvspace{8.0pt}%
\@x{ RsOK \.{\defeq}}%
\@x{\@s{16.4} \.{\land} RsTypeOK}%
\@x{\@s{16.4} \.{\land} RsPodUnique}%
\@x{\@s{16.4} \.{\land} RsPodComplete}%
\@pvspace{8.0pt}%
 \@x{ RsUnchanged \.{\defeq} {\UNCHANGED} {\langle} DesiredReplicas ,\,
 CreatedPods ,\, MaxPodId {\rangle}}%
\@pvspace{8.0pt}%
\@x{}\midbar\@xx{}%
\@pvspace{8.0pt}%
\@x{ DoRsInit \.{\defeq}}%
\@x{\@s{16.4} \.{\land} DesiredReplicas \.{=} 0}%
\@x{\@s{16.4} \.{\land} MaxPodId \.{=} 0}%
\@x{\@s{16.4} \.{\land} CreatedPods \.{=} \{ \}}%
\@pvspace{8.0pt}%
\@x{ DoRsCrash \.{\defeq}}%
\@x{\@s{16.4} \.{\land} CreatedPods \.{'} \.{=} APIPods}%
\@x{\@s{16.4} \.{\land} DesiredReplicas \.{'} \.{=} Cardinality ( APIPods )}%
\@x{\@s{16.4} \.{\land} {\UNCHANGED} MaxPodId}%
\@x{\@s{16.4} \.{\land} APIUnchanged}%
\@x{\@s{16.4} \.{\land} RsSchedDisconnected}%
\@x{\@s{16.4} \.{\land} AsRsDisconnected}%
\@x{\@s{16.4} \.{\land} {\UNCHANGED} {\langle} SchedKletConn {\rangle}}%
\@pvspace{8.0pt}%
\@x{ DoRsControlLoop \.{\defeq}}%
\@x{\@s{16.4} \.{\LET} tryRecv\@s{7.49} \.{\defeq}}%
\@x{\@s{53.19} {\IF} Len ( AsRsConn . inflight ) \.{>} 0 \.{\THEN}}%
 \@x{\@s{69.45} \.{\land} DesiredReplicas \.{'} \.{=} Head ( AsRsConn .
 inflight )}%
 \@x{\@s{69.45} \.{\land} AsRsConn \.{'} \.{=} [ AsRsConn {\EXCEPT} {\bang} .
 inflight \.{=} Tail ( @ ) ]}%
\@x{\@s{53.19} \.{\ELSE}}%
\@x{\@s{69.59} \.{\land} {\UNCHANGED} AsRsConn}%
\@x{\@s{69.59} \.{\land} {\UNCHANGED} DesiredReplicas}%
\@x{\@s{36.79} trySend ( batch ) \.{\defeq}}%
\@x{\@s{53.19} {\IF} RsSchedConn . connected \.{=} {\TRUE} \.{\THEN}}%
 \@x{\@s{69.45} RsSchedConn \.{'} \.{=} [ RsSchedConn {\EXCEPT} {\bang} .
 inflight \.{=} @ \.{\circ} {\langle} batch {\rangle} ]}%
\@x{\@s{53.19} \.{\ELSE} {\UNCHANGED} RsSchedConn}%
\@x{\@s{36.79} tryLoopOnce \.{\defeq}}%
 \@x{\@s{53.19} {\IF} DesiredReplicas \.{>} Cardinality ( CreatedPods )
 \.{\THEN}}%
 \@x{\@s{69.45} \.{\LET} newBatch \.{\defeq} [ batchId \.{\mapsto} MaxPodId
 ,\, batchSize \.{\mapsto} DesiredReplicas \.{-} Cardinality ( CreatedPods )
 ]}%
 \@x{\@s{89.85} newPods \.{\defeq} \{ newBatch . batchId \.{+} i \.{:} i
 \.{\in} 1 \.{\dotdot} newBatch . batchSize \} \.{\IN}}%
\@x{\@s{69.45} \.{\land} trySend ( newBatch )}%
 \@x{\@s{69.45} \.{\land} MaxPodId \.{'} \.{=} MaxPodId \.{+} newBatch .
 batchSize}%
 \@x{\@s{69.45} \.{\land} CreatedPods \.{'} \.{=} CreatedPods \.{\cup}
 newPods}%
\@x{\@s{53.19} \.{\ELSE}}%
 \@x{\@s{69.59} \.{\land} {\UNCHANGED} {\langle} CreatedPods ,\, MaxPodId
 {\rangle}}%
\@x{\@s{69.59} \.{\land} {\UNCHANGED} {\langle} RsSchedConn {\rangle}}%
\@x{\@s{16.4} \.{\IN}}%
\@x{\@s{16.4} \.{\land} tryRecv}%
\@x{\@s{16.4} \.{\land} tryLoopOnce}%
\@x{\@s{16.4} \.{\land} APIUnchanged}%
\@x{\@s{16.4} \.{\land} {\UNCHANGED} {\langle} SchedKletConn {\rangle}}%
\@pvspace{8.0pt}%
\@x{}\bottombar\@xx{}%
\@pvspace{8.0pt}%
\@x{}\moduleLeftDash\@xx{ {\MODULE} Scheduler}\moduleRightDash\@xx{}%
\@x{ {\EXTENDS} KdInterface}%
\@pvspace{8.0pt}%
\@x{ {\VARIABLES}}%
\@x{\@s{16.4} ScheduledPods}%
\@pvspace{8.0pt}%
\@x{ SchedTypeOK \.{\defeq}}%
 \@x{\@s{16.4} \.{\land} ScheduledPods \.{\in} {\SUBSET} [ pod \.{:} Nat ,\,
 node \.{:} Nat ]}%
 \@x{\@s{16.4} \.{\land} \A\, p \.{\in} ScheduledPods \.{:} p . node \.{\in} 1
 \.{\dotdot} NumNodes}%
\@pvspace{8.0pt}%
\@x{ SchedPodUnique \.{\defeq}}%
 \@x{\@s{16.4} \A\, p1 ,\, p2 \.{\in} ScheduledPods \.{:} p1 . pod \.{=} p2 .
 pod \.{\implies} p1 . node \.{=} p2 . node}%
\@pvspace{8.0pt}%
\@x{ SchedPodComplete \.{\defeq}}%
\@x{\@s{16.4} \A\, p \.{\in} APIPods \.{:} p \.{\in} ScheduledPods}%
\@pvspace{8.0pt}%
\@x{ SchedOK \.{\defeq}}%
\@x{\@s{16.4} \.{\land} SchedTypeOK}%
\@x{\@s{16.4} \.{\land} SchedPodUnique}%
\@x{\@s{16.4} \.{\land} SchedPodComplete}%
\@pvspace{8.0pt}%
 \@x{ SchedUnchanged \.{\defeq} {\UNCHANGED} {\langle} ScheduledPods
 {\rangle}}%
\@pvspace{8.0pt}%
\@x{}\midbar\@xx{}%
\@pvspace{8.0pt}%
\@x{ DoSchedInit \.{\defeq}}%
\@x{\@s{16.4} ScheduledPods \.{=} \{ \}}%
\@pvspace{8.0pt}%
\@x{ DoSchedCrash \.{\defeq}}%
\@x{\@s{16.4} \.{\land} ScheduledPods \.{'} \.{=} APIPods}%
\@x{\@s{16.4} \.{\land} APIUnchanged}%
\@x{\@s{16.4} \.{\land} SchedKletDisconnected}%
\@x{\@s{16.4} \.{\land} RsSchedDisconnected}%
\@x{\@s{16.4} \.{\land} {\UNCHANGED} {\langle} AsRsConn {\rangle}}%
\@pvspace{8.0pt}%
\@x{ DoSchedControlLoop \.{\defeq}}%
\@x{\@s{16.4} \.{\LET} trySend ( pods ) \.{\defeq}}%
\@x{\@s{53.19} {\IF} SchedKletConn . connected \.{=} {\TRUE} \.{\THEN}}%
 \@x{\@s{69.45} SchedKletConn \.{'} \.{=} [ SchedKletConn {\EXCEPT} {\bang} .
 inflight \.{=} @ \.{\circ} SetToSeq ( pods ) ]}%
\@x{\@s{53.19} \.{\ELSE} {\UNCHANGED} SchedKletConn}%
\@x{\@s{36.79} tryLoopOnce \.{\defeq}}%
\@x{\@s{53.19} {\IF} Len ( RsSchedConn . inflight ) \.{>} 0 \.{\THEN}}%
\@x{\@s{69.45} \.{\LET} newBatch \.{\defeq} Head ( RsSchedConn . inflight )}%
 \@x{\@s{89.85} newPods \.{\defeq} \{ [ pod \.{\mapsto} newBatch . batchId
 \.{+} i ,\, node \.{\mapsto} {\CHOOSE} n \.{\in} 1 \.{\dotdot} NumNodes
 \.{:} {\TRUE} ] \.{:} i \.{\in} 1 \.{\dotdot} newBatch . batchSize \}
 \.{\IN}}%
 \@x{\@s{69.45} \.{\land} RsSchedConn \.{'} \.{=} [ RsSchedConn {\EXCEPT}
 {\bang} . inflight \.{=} Tail ( @ ) ]}%
\@x{\@s{69.45} \.{\land} trySend ( newPods )}%
 \@x{\@s{69.45} \.{\land} ScheduledPods \.{'} \.{=} ScheduledPods \.{\cup} \{
 newPods \}}%
\@x{\@s{53.19} \.{\ELSE}}%
\@x{\@s{69.59} \.{\land} {\UNCHANGED} ScheduledPods}%
 \@x{\@s{69.59} \.{\land} {\UNCHANGED} {\langle} RsSchedConn ,\, SchedKletConn
 {\rangle}}%
\@x{\@s{16.4} \.{\IN}}%
\@x{\@s{16.4} \.{\land} tryLoopOnce}%
\@x{\@s{16.4} \.{\land} APIUnchanged}%
\@x{\@s{16.4} \.{\land} {\UNCHANGED} {\langle} AsRsConn {\rangle}}%
\@pvspace{8.0pt}%
\@x{}\bottombar\@xx{}%
\@pvspace{8.0pt}%
\@x{}\moduleLeftDash\@xx{ {\MODULE} Kubelet}\moduleRightDash\@xx{}%
\@x{ {\EXTENDS} KdInterface}%
\@pvspace{8.0pt}%
\@x{ {\VARIABLES}}%
\@x{\@s{16.4} RunningPods}%
\@pvspace{8.0pt}%
\@x{ KletTypeOK \.{\defeq}}%
 \@x{\@s{16.4} RunningPods \.{\in} {\SUBSET} [ pod \.{:} Nat ,\, node \.{:}
 Nat ]}%
\@pvspace{8.0pt}%
\@x{ KletPodUnique \.{\defeq}}%
 \@x{\@s{16.4} \A\, p1 ,\, p2 \.{\in} RunningPods \.{:} p1 . pod \.{=} p2 .
 pod \.{\implies} p1 . node \.{=} p2 . node}%
\@pvspace{8.0pt}%
\@x{ KletPodComplete \.{\defeq}}%
\@x{\@s{16.4} \A\, p \.{\in} APIPods \.{:} p \.{\in} RunningPods}%
\@pvspace{8.0pt}%
\@x{ KletOK \.{\defeq}}%
\@x{\@s{16.4} \.{\land}\@s{7.46} KletTypeOK}%
\@x{\@s{16.4} \.{\land}\@s{7.46} KletPodUnique}%
\@x{\@s{16.4} \.{\land}\@s{7.46} KletPodComplete}%
\@pvspace{8.0pt}%
\@x{ KletUnchanged \.{\defeq} {\UNCHANGED} {\langle} RunningPods {\rangle}}%
\@pvspace{8.0pt}%
\@x{}\midbar\@xx{}%
\@pvspace{8.0pt}%
\@x{ DoKletInit \.{\defeq}}%
\@x{\@s{16.4} RunningPods \.{=} \{ \}}%
\@pvspace{8.0pt}%
\@x{ DoKletCrash \.{\defeq}}%
\@x{\@s{16.4} \.{\land} RunningPods \.{'} \.{=} APIPods}%
\@x{\@s{16.4} \.{\land} APIUnchanged}%
\@x{\@s{16.4} \.{\land} SchedKletDisconnected}%
 \@x{\@s{16.4} \.{\land} {\UNCHANGED} {\langle} AsRsConn ,\, RsSchedConn
 {\rangle}}%
\@pvspace{8.0pt}%
\@x{ DoKletControlLoop \.{\defeq}}%
\@x{\@s{16.4} \.{\LET} tryExpose\@s{4.85} \.{\defeq}}%
 \@x{\@s{53.19} \.{\LET} pendingPods \.{\defeq} RunningPods \.{\,\backslash\,}
 APIPods \.{\IN}}%
 \@x{\@s{53.19} \.{\land} APIPods \.{'} \.{=} APIPods \.{\cup} \{ {\CHOOSE} p
 \.{\in} pendingPods \.{:} {\TRUE} \}}%
\@x{\@s{53.19} \.{\land} {\UNCHANGED} CmdIndex}%
\@x{\@s{36.79} tryLoopOnce \.{\defeq}}%
\@x{\@s{53.19} {\IF} Len ( SchedKletConn . inflight ) \.{>} 0 \.{\THEN}}%
 \@x{\@s{69.45} \.{\LET} newPod \.{\defeq} Head ( SchedKletConn . inflight )
 \.{\IN}}%
 \@x{\@s{69.45} \.{\land} SchedKletConn \.{'} \.{=} [ SchedKletConn {\EXCEPT}
 {\bang} . inflight \.{=} Tail ( @ ) ]}%
 \@x{\@s{69.45} \.{\land} RunningPods \.{'} \.{=} RunningPods \.{\cup} \{
 newPod \}}%
\@x{\@s{53.19} \.{\ELSE}}%
\@x{\@s{69.59} \.{\land} {\UNCHANGED} RunningPods}%
\@x{\@s{69.59} \.{\land} {\UNCHANGED} SchedKletConn}%
\@x{\@s{16.4} \.{\IN}}%
\@x{\@s{16.4} \.{\land} tryLoopOnce}%
\@x{\@s{16.4} \.{\land} tryExpose}%
 \@x{\@s{16.4} \.{\land} {\UNCHANGED} {\langle} AsRsConn ,\, RsSchedConn
 {\rangle}}%
\@pvspace{8.0pt}%
\@x{}\bottombar\@xx{}%
\@pvspace{8.0pt}%
\@x{}\moduleLeftDash\@xx{ {\MODULE} Kubedirect}\moduleRightDash\@xx{}%
 \@x{ {\EXTENDS} KdInterface ,\, Autoscaler ,\, ReplicaSet ,\, Scheduler ,\,
 Kubelet}%
\@pvspace{8.0pt}%
\@x{ KdVars \.{\defeq} {\langle} CmdIndex ,\, APIPods ,\,}%
\@x{\@s{56.48} AsRsConn ,\, RsSchedConn ,\, SchedKletConn ,\,}%
\@x{\@s{56.48} LastDesiredReplicas ,\,}%
\@x{\@s{56.48} DesiredReplicas ,\, CreatedPods ,\, MaxPodId ,\,}%
\@x{\@s{56.48} ScheduledPods ,\,}%
\@x{\@s{56.48} RunningPods}%
\@x{\@s{56.48} {\rangle}}%
\@pvspace{8.0pt}%
\@x{ KdDoScalingCmd \.{\defeq}}%
\@x{\@s{16.4} \.{\land} {\IF} CmdIndex \.{<} Len ( ScalingCmds ) \.{\THEN}}%
\@x{\@s{43.76} CmdIndex \.{'} \.{=} CmdIndex \.{+} 1}%
\@x{\@s{27.51} \.{\ELSE}}%
\@x{\@s{43.91} {\UNCHANGED} CmdIndex}%
\@x{\@s{16.4} \.{\land} {\UNCHANGED} APIPods}%
 \@x{\@s{16.4} \.{\land} {\UNCHANGED} {\langle} AsRsConn ,\, RsSchedConn ,\,
 SchedKletConn {\rangle}}%
\@x{\@s{16.4} \.{\land} AsUnchanged}%
\@x{\@s{16.4} \.{\land} RsUnchanged}%
\@x{\@s{16.4} \.{\land} SchedUnchanged}%
\@x{\@s{16.4} \.{\land} KletUnchanged}%
\@pvspace{8.0pt}%
\@x{ KdDoAsNext \.{\defeq}}%
\@x{\@s{16.4} \.{\land} DoAsControlLoop}%
\@x{\@s{16.4} \.{\land} RsUnchanged}%
\@x{\@s{16.4} \.{\land} SchedUnchanged}%
\@x{\@s{16.4} \.{\land} KletUnchanged}%
\@pvspace{8.0pt}%
\@x{ KdDoRsNext \.{\defeq}}%
\@x{\@s{16.4} \.{\land} DoRsControlLoop}%
\@x{\@s{16.4} \.{\land} AsUnchanged}%
\@x{\@s{16.4} \.{\land} SchedUnchanged}%
\@x{\@s{16.4} \.{\land} KletUnchanged}%
\@pvspace{8.0pt}%
\@x{ KdDoSchedNext \.{\defeq}}%
\@x{\@s{16.4} \.{\land} DoSchedControlLoop}%
\@x{\@s{16.4} \.{\land} AsUnchanged}%
\@x{\@s{16.4} \.{\land} RsUnchanged}%
\@x{\@s{16.4} \.{\land} KletUnchanged}%
\@pvspace{8.0pt}%
\@x{ KdDoKletNext \.{\defeq}}%
\@x{\@s{16.4} \.{\land} DoKletControlLoop}%
\@x{\@s{16.4} \.{\land} AsUnchanged}%
\@x{\@s{16.4} \.{\land} RsUnchanged}%
\@x{\@s{16.4} \.{\land} SchedUnchanged}%
\@pvspace{8.0pt}%
\@x{}\midbar\@xx{}%
\@pvspace{8.0pt}%
\@x{ KdDoAsRsDisconnect \.{\defeq}}%
\@x{\@s{16.4} \.{\land} AsRsDisconnected}%
 \@x{\@s{16.4} \.{\land} {\UNCHANGED} {\langle} RsSchedConn ,\, SchedKletConn
 {\rangle}}%
\@x{\@s{16.4} \.{\land} APIUnchanged}%
\@x{\@s{16.4} \.{\land} AsUnchanged}%
\@x{\@s{16.4} \.{\land} RsUnchanged}%
\@x{\@s{16.4} \.{\land} SchedUnchanged}%
\@x{\@s{16.4} \.{\land} KletUnchanged}%
\@pvspace{8.0pt}%
\@x{ KdDoRsSchedDisconnect \.{\defeq}}%
\@x{\@s{16.4} \.{\land} RsSchedDisconnected}%
 \@x{\@s{16.4} \.{\land} {\UNCHANGED} {\langle} AsRsConn ,\, SchedKletConn
 {\rangle}}%
\@x{\@s{16.4} \.{\land} APIUnchanged}%
\@x{\@s{16.4} \.{\land} AsUnchanged}%
\@x{\@s{16.4} \.{\land} RsUnchanged}%
\@x{\@s{16.4} \.{\land} SchedUnchanged}%
\@x{\@s{16.4} \.{\land} KletUnchanged}%
\@pvspace{8.0pt}%
\@x{ KdDoSchedKletDisconnect \.{\defeq}}%
\@x{\@s{16.4} \.{\land} SchedKletDisconnected}%
 \@x{\@s{16.4} \.{\land} {\UNCHANGED} {\langle} AsRsConn ,\, RsSchedConn
 {\rangle}}%
\@x{\@s{16.4} \.{\land} APIUnchanged}%
\@x{\@s{16.4} \.{\land} AsUnchanged}%
\@x{\@s{16.4} \.{\land} RsUnchanged}%
\@x{\@s{16.4} \.{\land} SchedUnchanged}%
\@x{\@s{16.4} \.{\land} KletUnchanged}%
\@pvspace{8.0pt}%
\@x{ KdDoAsCrash \.{\defeq}}%
\@x{\@s{16.4} \.{\land} DoAsCrash}%
\@x{\@s{16.4} \.{\land} RsUnchanged}%
\@x{\@s{16.4} \.{\land} SchedUnchanged}%
\@x{\@s{16.4} \.{\land} KletUnchanged}%
\@pvspace{8.0pt}%
\@x{ KdDoRsCrash \.{\defeq}}%
\@x{\@s{16.4} \.{\land} DoRsCrash}%
\@x{\@s{16.4} \.{\land} AsUnchanged}%
\@x{\@s{16.4} \.{\land} SchedUnchanged}%
\@x{\@s{16.4} \.{\land} KletUnchanged}%
\@pvspace{8.0pt}%
\@x{ KdDoSchedCrash \.{\defeq}}%
\@x{\@s{16.4} \.{\land} DoSchedCrash}%
\@x{\@s{16.4} \.{\land} AsUnchanged}%
\@x{\@s{16.4} \.{\land} RsUnchanged}%
\@x{\@s{16.4} \.{\land} KletUnchanged}%
\@pvspace{8.0pt}%
\@x{ KdDoKletCrash \.{\defeq}}%
\@x{\@s{16.4} \.{\land} DoKletCrash}%
\@x{\@s{16.4} \.{\land} AsUnchanged}%
\@x{\@s{16.4} \.{\land} RsUnchanged}%
\@x{\@s{16.4} \.{\land} SchedUnchanged}%
\@pvspace{8.0pt}%
\@x{ KdDoSchedKletHandshake \.{\defeq}}%
\@x{\@s{16.4} \.{\LET} tryHandshake \.{\defeq}}%
 \@x{\@s{36.79} \.{\land} {\IF} SchedKletConn . connected \.{=} {\FALSE}
 \.{\THEN}}%
 \@x{\@s{64.16} \.{\LET} allPods \.{\defeq} \{ RunningPods \.{\cup}
 ScheduledPods \} \.{\IN}}%
\@x{\@s{64.16} \.{\land} ScheduledPods \.{'}\@s{3.91} \.{=} allPods}%
 \@x{\@s{64.16} \.{\land} SchedKletConn \.{'} \.{=} [ SchedKletConn {\EXCEPT}
 {\bang} . inflight \.{=} SetToSeq ( allPods \.{\,\backslash\,} RunningPods )
 ,\,}%
\@x{\@s{270.17} {\bang} . connected \.{=} {\TRUE} ]}%
\@x{\@s{47.91} \.{\ELSE}}%
\@x{\@s{64.31} \.{\land} {\UNCHANGED} ScheduledPods}%
\@x{\@s{64.31} \.{\land} {\UNCHANGED} SchedKletConn}%
 \@x{\@s{36.79} \.{\land} {\UNCHANGED} {\langle} AsRsConn ,\, RsSchedConn
 {\rangle}}%
\@x{\@s{36.79} \.{\land} APIUnchanged}%
\@x{\@s{16.4} \.{\IN}}%
\@x{\@s{16.4} \.{\land} tryHandshake}%
\@x{\@s{16.4} \.{\land} AsUnchanged}%
\@x{\@s{16.4} \.{\land} RsUnchanged}%
\@x{\@s{16.4}}%
\@y{%
  /\ SchedUnchanged
}%
\@xx{}%
\@x{\@s{16.4} \.{\land} KletUnchanged}%
\@pvspace{16.0pt}%
\@x{ KdDoRsSchedHandshake \.{\defeq}}%
\@x{\@s{16.4} \.{\LET} tryHandshake\@s{12.18} \.{\defeq}}%
 \@x{\@s{36.79} \.{\land} {\IF} RsSchedConn . connected \.{=} {\FALSE}
 \.{\land} SchedKletConn . connected \.{=} {\TRUE} \.{\THEN}}%
\@x{\@s{64.16} \.{\land} CreatedPods \.{'} \.{=} ScheduledPods}%
 \@x{\@s{64.16} \.{\land} DesiredReplicas \.{'} \.{=} Cardinality (
 ScheduledPods )}%
 \@x{\@s{64.16} \.{\land} RsSchedConn \.{'} \.{=} [ RsSchedConn {\EXCEPT}
 {\bang} . connected \.{=} {\TRUE} ]}%
\@x{\@s{47.91} \.{\ELSE}}%
 \@x{\@s{64.31} \.{\land} {\UNCHANGED} {\langle} CreatedPods ,\,
 DesiredReplicas {\rangle}}%
\@x{\@s{64.31} \.{\land} {\UNCHANGED} RsSchedConn}%
\@x{\@s{36.79} \.{\land} {\UNCHANGED} {\langle} MaxPodId {\rangle}}%
 \@x{\@s{36.79} \.{\land} {\UNCHANGED} {\langle} AsRsConn ,\, SchedKletConn
 {\rangle}}%
\@x{\@s{36.79} \.{\land} APIUnchanged}%
\@x{\@s{16.4} \.{\IN}}%
\@x{\@s{16.4} \.{\land} tryHandshake}%
\@x{\@s{16.4} \.{\land} AsUnchanged}%
\@x{\@s{16.4}}%
\@y{%
  /\ RsUnchanged
}%
\@xx{}%
\@x{\@s{16.4} \.{\land} SchedUnchanged}%
\@x{\@s{16.4} \.{\land} KletUnchanged}%
\@pvspace{8.0pt}%
\@x{ KdDoAsRsHandshake \.{\defeq}}%
\@x{\@s{16.4} \.{\LET} tryHandshake \.{\defeq}}%
 \@x{\@s{36.79} \.{\land} {\IF} AsRsConn . connected \.{=} {\FALSE} \.{\land}
 RsSchedConn . connected \.{=} {\TRUE} \.{\THEN}}%
 \@x{\@s{64.16} AsRsConn \.{'} \.{=} [ AsRsConn {\EXCEPT} {\bang} . connected
 \.{=} {\TRUE} ]}%
\@x{\@s{47.91} \.{\ELSE}}%
\@x{\@s{64.31} {\UNCHANGED} AsRsConn}%
 \@x{\@s{36.79} \.{\land} {\UNCHANGED} {\langle} RsSchedConn ,\, SchedKletConn
 {\rangle}}%
\@x{\@s{36.79} \.{\land} APIUnchanged}%
\@x{\@s{16.4} \.{\IN}}%
\@x{\@s{16.4} \.{\land} tryHandshake}%
\@x{\@s{16.4} \.{\land} AsUnchanged}%
\@x{\@s{16.4} \.{\land} RsUnchanged}%
\@x{\@s{16.4} \.{\land} SchedUnchanged}%
\@x{\@s{16.4} \.{\land} KletUnchanged}%
\@pvspace{8.0pt}%
\@x{}\midbar\@xx{}%
\@pvspace{8.0pt}%
\@x{ KdInit \.{\defeq}}%
\@x{\@s{16.4} \.{\land}\@s{2.10} DoAPIInit}%
\@x{\@s{16.4} \.{\land}\@s{2.10} DoConnInit}%
\@x{\@s{16.4} \.{\land}\@s{2.10} DoAsInit}%
\@x{\@s{16.4} \.{\land}\@s{2.10} DoRsInit}%
\@x{\@s{16.4} \.{\land}\@s{2.10} DoSchedInit}%
\@x{\@s{16.4} \.{\land}\@s{2.10} DoKletInit}%
\@pvspace{8.0pt}%
\@x{ KdNext \.{\defeq}}%
\@x{\@s{16.4} \.{\lor}\@s{6.23} KdDoScalingCmd}%
\@x{\@s{16.4} \.{\lor}\@s{6.23} KdDoAsNext}%
\@x{\@s{16.4} \.{\lor}\@s{6.23} KdDoRsNext}%
\@x{\@s{16.4} \.{\lor}\@s{6.23} KdDoSchedNext}%
\@x{\@s{16.4} \.{\lor}\@s{6.23} KdDoKletNext}%
\@x{\@s{16.4} \.{\lor}\@s{6.23} KdDoAsRsDisconnect}%
\@x{\@s{16.4} \.{\lor}\@s{6.23} KdDoRsSchedDisconnect}%
\@x{\@s{16.4} \.{\lor}\@s{6.23} KdDoSchedKletDisconnect}%
\@x{\@s{16.4} \.{\lor}\@s{6.23} KdDoAsCrash}%
\@x{\@s{16.4} \.{\lor}\@s{6.23} KdDoRsCrash}%
\@x{\@s{16.4} \.{\lor}\@s{6.23} KdDoSchedCrash}%
\@x{\@s{16.4} \.{\lor}\@s{6.23} KdDoKletCrash}%
\@x{\@s{16.4} \.{\lor}\@s{6.23} KdDoSchedKletHandshake}%
\@x{\@s{16.4} \.{\lor}\@s{6.23} KdDoRsSchedHandshake}%
\@x{\@s{16.4} \.{\lor}\@s{6.23} KdDoAsRsHandshake}%
\@pvspace{8.0pt}%
\@x{ KdInvariants \.{\defeq}}%
\@x{\@s{16.4} \.{\land} InterfaceOK}%
\@x{\@s{16.4} \.{\land} AsOK}%
\@x{\@s{16.4} \.{\land} RsOK}%
\@x{\@s{16.4} \.{\land} SchedOK}%
\@x{\@s{16.4} \.{\land} KletOK}%
\@pvspace{8.0pt}%
\@x{ KdConvergence \.{\defeq}}%
\@x{\@s{16.4} Cardinality ( APIPods ) \.{=} ScalingCmds [ CmdIndex ]}%
\@pvspace{8.0pt}%
\@x{ KdFairness \.{\defeq}}%
\@x{\@s{16.4} \.{\land} {\WF}_{ KdVars} ( KdDoAsNext )}%
\@x{\@s{16.4} \.{\land} {\WF}_{ KdVars} ( KdDoRsNext )}%
\@x{\@s{16.4} \.{\land} {\WF}_{ KdVars} ( KdDoSchedNext )}%
\@x{\@s{16.4} \.{\land} {\WF}_{ KdVars} ( KdDoKletNext )}%
\@x{\@s{16.4} \.{\land} {\WF}_{ KdVars} ( KdDoAsRsHandshake )}%
\@x{\@s{16.4} \.{\land} {\WF}_{ KdVars} ( KdDoRsSchedHandshake )}%
\@x{\@s{16.4} \.{\land} {\WF}_{ KdVars} ( KdDoSchedKletHandshake )}%
\@pvspace{8.0pt}%
\@x{ KdSafety \.{\defeq} {\Box} KdInvariants}%
\@pvspace{8.0pt}%
 \@x{ KdLiveness \.{\defeq} {\Diamond} {\Box} KdConvergence \.{\land}
 KdFairness}%
\@pvspace{8.0pt}%
 \@x{ Spec \.{\defeq} KdInit \.{\land} {\Box} [ KdNext ]_{ KdVars} \.{\land}
 KdSafety \.{\land} KdLiveness}%
\@pvspace{8.0pt}%
\@x{}\bottombar\@xx{}%
\@pvspace{8.0pt}%
\end{tlatex}

\end{document}